\documentclass[a4paper, 11pt]{article}

\pdfoutput=1

\usepackage{graphicx}
\usepackage{amsmath}
\usepackage{amsfonts}
\usepackage{bm}
\usepackage{a4wide}
\usepackage{natbib}

\newcommand\eq[1]{(\ref{#1})}
\newcommand\pd{\partial}
\newcommand\bmi{\bm}
\newcommand\Kb{{\bmi{K}}}
\newcommand\wb{{\bmi{w}}}
\newcommand\nb{{\bmi{n}}}
\newcommand\vb{{\bmi{v}}}
\newcommand\ie{\textit{i.e.}~}

\begin{document}

\title{Modeling of the contrast-enhanced perfusion test in liver based on the
multi-compartment flow in porous media}

%
\author{%
  Eduard Rohan,
  Vladim\'ir Luke\v{s},
  Alena Jon\'a\v{s}ov\'a\medskip\\
  {\it\small New Technologies for the Information Society,}\\
  {\it\small Faculty of Applied Sciences, University of West Bohemia,}\\
  {\it\small Univerzitn\'{\i}~8, 30614, Pilsen, Czech Republic,}\\
  {\it\small e-mail: rohan@kme.zcu.cz}
  }
%

%
\date{May, 2016}

\def\keywords#1{\noindent{\bf Keywords:} #1}
\def\abstract#1{\noindent{\bf Abstract:} #1}
\def\acknowledgements#1{\noindent{\bf Acknowledgements:} #1}

\maketitle

%
\abstract{%
The paper deals with modeling the liver perfusion intended to improve
quantitative analysis of the tissue scans provided by the contrast-enhanced computed tomography
(CT). For this purpose, we developed a model of dynamic transport of the
contrast fluid through the hierarchies of the perfusion trees. Conceptually,
computed time-space distributions of the so-called tissue density can be
compared with the measured data obtained from CT; such a modeling feedback can
be used for model parameter identification. The blood flow is characterized at
several scales for which different models are used. Flows in upper hierarchies
represented by larger branching vessels are described using simple 1D models
based on the Bernoulli equation extended by correction terms to respect the
local pressure losses. To describe flows in smaller vessels and in the tissue
parenchyma, we propose a 3D continuum model of porous medium defined in terms of
hierarchically matched compartments characterized by hydraulic permeabilities.
The 1D models corresponding to the portal and hepatic veins are coupled with the
3D model through point sources, or sinks. The contrast fluid saturation is
governed by transport equations adapted for the 1D and 3D flow models.  The
complex perfusion model has been implemented using the finite element and finite
volume methods.  We report numerical examples computed for anatomically relevant
geometries of the liver organ and of the principal vascular trees. The simulated
tissue density corresponding to the CT examination output reflects a pathology
modeled as a localized permeability deficiency.}


\newpage

\section{Introduction}\label{sec-intro}

Modeling blood perfusion in tissues belongs to the most challenging
problems in the biomechanical and biomedical research. The topic is of
interest especially in liver surgery
\citep{patient_specific_hepatic-2016}, brain neurosurgery, or neurology
\citep{voxelized_brain-2016,vascular_grap-2009,cerebral_flow-2011}. Although
the functionality of the two related organs, liver and brain, is
completely different, there are some similar aspects for both. In
contrast with the myocardium, another highly perfused tissue, for both
the organs the phenomenon of deformation can be neglected unless
special concerns are in focus, see {e.g.{~}}
\cite{vascularized_liver-2012}. It is desirable to find accurately
parts of the tissue with an insufficient blood supply, to localize
anomalies in the blood microcirculation, and to quantify locally the
perfusion efficiency. It is worth to note that also the kidney belongs
to highly perfused tissues with its important role in the blood
circulation system, but we do not tackle this issue in the present
paper.

In most perfused organs, the blood is transported through complex hierarchical
branching systems of vessels with various diameters which diminish consecutively
at each bifurcation, beginning from arteries down to capillaries. The ascending
system of veins is arranged in the reversed sense, starting at the capillary
level, where the blood participates in the metabolism process,
delivering the oxygen to the tissue and collecting the waste products. The blood
circulation in the tissue can be represented by two penetrating, but mutually
separated {vascular} trees connected only at the capillary level.  In reality,
collateralization vessels exist which make shunts between the two trees, but
this phenomenon is usually negligible for the global hierarchically built
system.

Although the modeling approach is presented in a rather general
setting, in this paper, we focus on the \emph{liver tissue} which has
a special structure. The blood flows in the whole organ from two
separated vascular trees, one connected to the hepatic artery (HA) and
the other to the portal vein (PV), whereby the venous supply
dominates. The two branching systems bifurcate to about 20
generations \citep{Debbaut-JOA2014}. The blood is filtered in the
tissue functional units, called hepatic units, or lobules, which
typically are considered as hexagonal prisms. At their corner edges,
the hepatic artery and portal vein trees terminate and the blood flows
through hepatic capillaries constituting the sinusoids to the
so-called central veins, the terminal branches of the draining hepatic
vein (HV) located in the center of each lobule. In the proposed model,
for simplicity we neglect the HA, as its blood supply is minor in the
comparison with the one by the portal vein\footnote{The perfusion
  model is intended to estimate the volume regeneration capacity of
  the liver parenchyma \citep{bruha_2015}. In this context, it is
  important to capture the supply through the portal vein.  Concerning
  the perfusion CT simulation, we focus on the second stage when the
  contrast bolus arrives at the PV inlet so that the first stage
  associated with the HA is over.}.

The steadily growing body of published works related to the modeling of liver
perfusion shows its importance for the medical research. Obviously, relevant
computational models which can be used for complex simulations involving various
parameters require detailed information about the liver structure.  However,
there is a pertaining difficulty in describing the vascular trees bridging
several scales. The current standard computed tomography (CT), magnetic
resonance imaging (MRI) and related image processing techniques provide
resolution of the vascular trees down to millimeters, which is not enough to
capture details on about 10 underlying tree generations. Therefore, various
methods and algorithms have been proposed to generate artificial, but
anatomically correct vascular architectures. These are coupled with the
authentic patient-specific reconstructed geometry to obtain complete vascular
trees with a required resolution of microvessels. The most popular approach is
based on the constrained constructive optimization (CCO), see e.g.~\cite{gco},
\cite{Schwen2012}. However, there are more advanced approaches based on the metabolism driven processes of
the vascular tree reconstruction
\citep{Schneider2012}.

Rather than modeling the flow on the complex branching system, many publications
have been devoted to detailed studies of microcirclation. Perfusion
characteristics of blood flow through hepatic microcapillaries of human lobular
sinusoids were analyzed by \cite{Debbaut2012} using tissue samples processed by
corrosion casting and the micro-CT imaging; anisotropic tensors of the
sinusoidal permeability were calculated by a simple averaging of the flow field
obtained using standard CFD simulations. Blood circulation in the liver {lobule}
was considered in \cite{Bonfiglio2010} using both Newtonian and shear-thinning
flow models. The liver tissue is able to adapt rapidly to modified hydraulic
conditions; in \cite{Ricken2010}, the sinusoidal permeability has been studied
in terms of a vascular remodeling process induced by outflow obstruction.
Sinusoidal vascular canals were assumed to align with the blood pressure
gradient. In \cite{Ricken-Werner-BMMB2015}, using the general framework of a
two-scale approach and the theory of porous media, the basic coupled
transport-reaction mechanisms between blood perfusion and the hepatic cell
metabolism were described at the level of lobule.
Lymphatic flow production in the liver lobules arranged in periodic lattices was
treated in \cite{Siggers2014}. A similar approach has been used in
\cite{Lettmann2014} to study microcirculation driven transport of oxygen and
cytokine in the tissue. Besides the blood flow in healthy liver itself,
diseased tissue deserve attention from medical point of view, but also as a
challenge for mathematical modeling; there is a considerable number of issues
related to modeling pathophysiological conditions, such as the fluid transport
through vascular tumors \citep{pozrikidis_2010,Shipley-Chapman2010,Mescam-TMI2010}.

\begin{figure}
  \centering
  \includegraphics[width=0.9\linewidth]{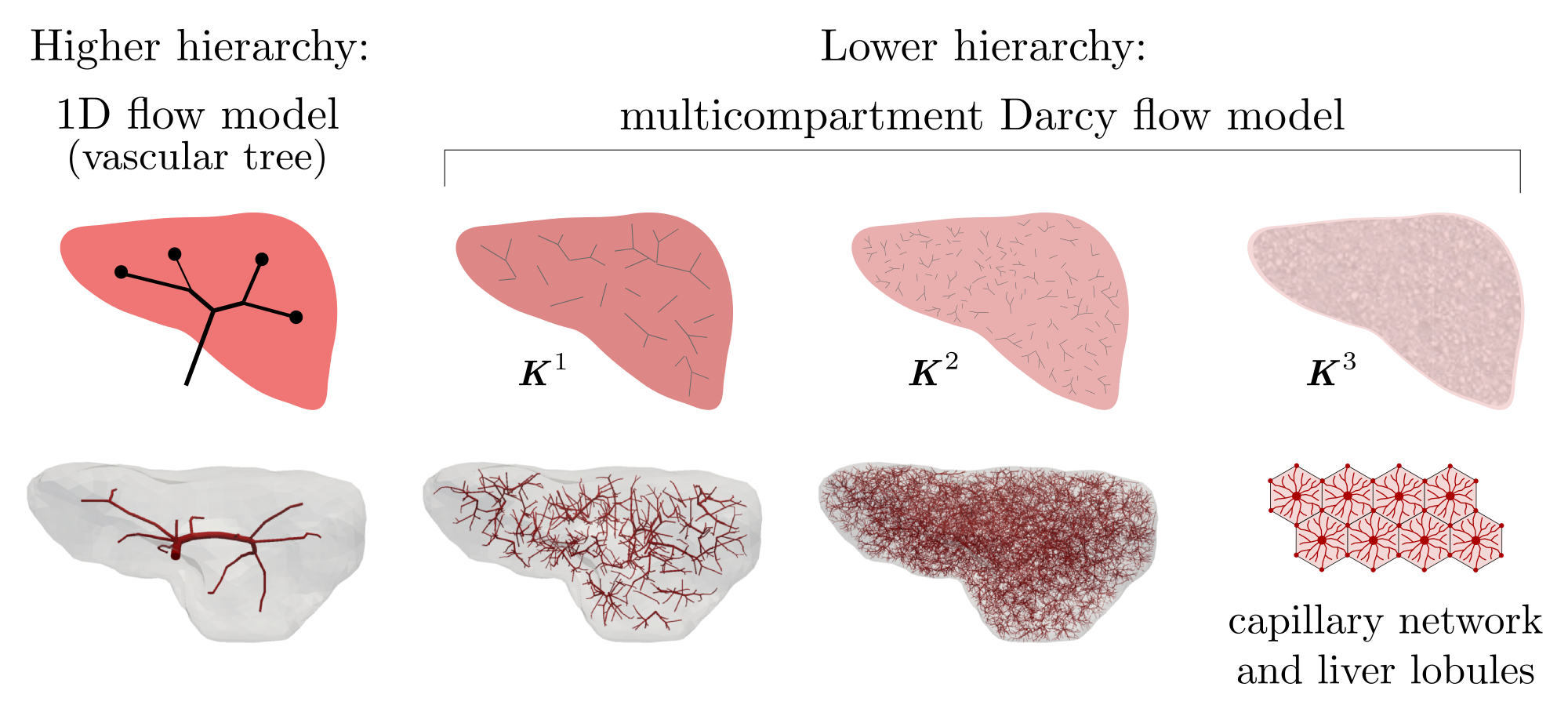}
  \caption{
    Decomposition of the blood vessel perfusion system in the liver:
    higher hierarchy (HH) -- blood flow described by a 1D model,
    lower hierarchy (LH) -- blood distribution governed by the Darcy flow
    model extended for multiple compartments, each compartment characterized by
    a permeability tensor $\Kb^{i}$}
  \label{fig-multiperf}
\end{figure}

Since the straightforward mathematical description of blood flow on the complete
vascular trees featured by several hierarchies using the Navier-Stokes equations
leads to a prohibitively complicated numerical model, some convenient model
reduction is needed. We adhere to the general approach proposed in
\cite{FormQuart-book} which consists in decomposing the tree into a higher (HH)
and a lower (LH) hierarchy,  see Fig.~\ref{fig-multiperf}.

The HH tree is constituted by larger vessels, so that the flow can be described
in the 3D geometry using the Navier-Stokes equations. Alternatively the vessels
can be replaced with line segments, so that the flow model is obtained using the
1D description of flow on such branching system of line segments.

To describe the blood flow in the LH tree, we resort to a type of the Darcy flow
model which is defined in the 3D bulk of the liver parenchyma. Since the LH part
involves vessels associated with different hierarchies, several compartments are
introduced and Darcy model is used within each of them. The flow between
different compartments is described in terms of the difference between the
compartmental pressure at a given location; details are given below.

\paragraph{Remark}\label{rem1}
The \textbf{multicompartment model of perfusion} has been used {e.g.{~}} in
\cite{Cimrman2007} and then in \cite{Rohan-Jonasova-Lukes-WCCM2014-liver} as a
phenomenological model. Recently a multicompartment model of cardiac perfusion
was presented in \cite{Michler2012}, cf. \cite{Michler2013}. For two
compartments it can be derived using the homogenization of the Darcy flow in
porous medium with large contrasts in the permeability
\citep{rohan-cimrman-perfusionIJMCE2010,Rohan-Lukes-Jonasova-ECCOMAS2012-layer-perf}.
Another derivation is based on the averaging theory by Slattery and Whitaker
\citep{Slattery1967,Whitaker1967} employed to the hierarchical flow in pores
with continuously variable cross-sections, cf.
\cite{huyghe_campen_1995:ALL}.\bigskip

In the context of the multi-porous, media employed in the geophysical research,
phenomenological approaches based on extensions of the Biot model
\citep{Mehrabian-Abousleiman-JGR2014} lead to multi-compartment models where
co-existing porosities are associated with distinct pressures. As a special
case, the double-porous, or double-permeability media was obtained by
homogenization approach applied to upscale periodic structures with large
contrasts in the hydraulic permeability
\citep{rohan-cimrman-perfusionIJMCE2010,rohan-etal-jmps2012-bone}. The derived
model describing fluid flow in the double-permeability medium with two systems
of periodic channels separated by low permeable matrix can be adapted to the
liver parenchyma generated as a quasi-periodic structure of lobules.

The aim of the present paper is twofold:
\begin{enumerate}
\item
we present a  numerically tractable model of the liver perfusion intended for real-time simulations, which is
based on the hierarchical decomposition of the perfusion trees, combining approaches of the multicompartment porous media flow and the  ``1D'' flow on vascular trees;
\item
we propose an associated model of transport of the contrast agent dissolved in the blood
and characterized by the saturation function.
\end{enumerate}
The latter model provides an analogous output to the so-called
\emph{tissue density} (measured in the Hounsfield units) which is
obtained from the medical perfusion examination by the dynamic contrast-enhanced CT, or alternatively by the magnetic resonance imaging, see
{e.g.{~}} \cite{fieselmann2011deconvolution,keeling_2007}.
We would like to stress out that this modeling option is important for the model parameters identification and, thereby, also for
the model validation.

The conception of the
hierarchical modeling reported in this paper has been outlined in a
short proceedings paper
\citep{Rohan-Jonasova-Lukes-WCCM2014-liver}. However, here we give
additional details on the model itself and present new results
computed for anatomically relevant data, including the anisotropic
permeability coefficients estimated by an analysis of the perfusion
trees.

The organization of the paper is as follows. The perfusion model based on the
hierarchical decomposition of the perfusion trees (the portal and hepatic veins)
is described in Section~\ref{sec-FM}. In Section~\ref{sec-CT}, the transport of
the contrast fluid is introduced. In Section~\ref{sec-simulation} we present
numerical examples to illustrate an application of the model to describe the
hierarchical blood flow in a ``liver'' which is a simplified geometrical
structure created using anatomical data combined with the CCO approach to define
the portal and hepatic vascular trees. The paper conclusions and the discussion
of further perspectives is presented in Section~\ref{sec-Concl}.

\section{Hierarchical mathematical model of perfusion}\label{sec-FM}

The simulation of blood perfusion in tissues, such as the hepatic or cerebral
ones, belongs to problems typically requiring a sort of multiscale modeling
approach \citep{DAngelo-thesis}. The term ``multiscale'' is employed in a
slightly different manner than in \cite{FormQuart-book}, where the ``geometrical
multiscale'' modeling was introduced in the context of the cardiovascular
system. The difficulty of blood perfusion modeling arises from the nature of
the flow in the perfusion trees incorporating blood vessels of very different
diameters. In liver, the blood is conveyed to the organ by the portal vein which
has the diameter of about $D=1\,\textrm{cm}$. The perfusion trees have several
hierarchies associated with bifurcations, where vessel diameters reduce. The
tissue parenchyma is arranged in hexagonal structures (lobules) with the
characteristic size of 1.5 mm where the blood is transported through the
capillaries with the diameter of about $d=10\,\mu\textrm{m}$ from the portal
compartment to the hepatic one. Thus, the overall scale change characterized by
the ratio of the largest and smallest diameters of the vascular trees is
approximately $d/D = 0.001$.

The model which we develop should allow also for simulation of the transport of
the contrast fluid (the tracer) during the contrast-enhanced dynamic perfusion test. The aim is to
provide a computational feedback which would enable us to analyze more
accurately the CT scans obtained from the standard dynamic perfusion test of a
patient \citep{Materne-CT-liver-perfus2000}. The methods currently being
used are based on the deconvolution, or maximum slope techniques. They provide
some local integral characteristics like the blood flow, blood volume, mean
transition times and others for each voxel of the tissue
\citep{Koh-etal-2006,fieselmann2011deconvolution}. Our approach should provide an
alternative and more detailed interpretation of the measured CT scans. The
proposed strategy announced first in our conference paper
\citep{Rohan-Jonasova-Lukes-WCCM2014-liver} is based on the following tasks:
\begin{itemize}
  \item to describe flows in  the perfusion trees including the parenchyma by
  combining suitable models which are associated with different hierarchies and
  which use the most information about the structures and geometry, involving
  only few undetermined parameters;
  \item to reconstruct the organ (liver) shape and the blood vessels geometries
  up to a certain hierarchy using the image segmentation techniques (based on
  the CT ``static'' data) \citep{Jirik2013,LISA} possibly supplemented by the
  CCO approach employed also in \cite{euroscipy2014};
  \item to identify selected parameters of the perfusion model (like
  permeability), so that the simulated dynamic CT examination well approximates
  the measured data \citep{Rohan-vipimage2015}.
\end{itemize}

Such a ``tuned'' model would enable us to analyze the blood flow in particular
compartments and to predict effects of intended medical treatment, like
resection of a part of the tissue.

In the proposed strategy, several difficulties arise: on one hand, the model
should reflect the microstructure and fit the complex geometry of the perfusion
trees, on the other hand, it should be parametrized using just not too many
parameters, to prevent ill conditioning of the ``tuning'' step {involving}
simulations of the steady blood flow and the dynamic tracer transport.
Therefore, the tuning parameters must reflect some important features of the
perfusion system.

The model of the perfusion and of the contrast fluid transport described in this
paper consists of the following two parts:
\begin{enumerate}
  \item The ``inlet'' and ``outlet'' trees which are associated with the portal
  and hepatic vein networks\footnote{We have in mind the liver perfusion
  modeling, so that we also adapt the notations to fit with this particular
  application.} (in the case of the liver; recall that we have neglected the
  hepatic artery in the model). These trees are denoted by $\mathcal{T}_P$ and
  $\mathcal{T}_H$ (label $P$ stands for the portal vein system, whereas label
  $H$ stands for the hepatic vein system). They are formed by vessel segments
  and junctions representing the bifurcations. The steady flow on these trees is
  described by a 1D model which is introduced in Section~\ref{sec-Bf}, see also
  \cite{Jonasova-1D-ACM-2014}. The model is relevant for higher hierarchies of
  the vascular trees, down to a certain size of the vessels, so that the
  terminal branches\footnote{or leaves in the graph theory terminology} of
  $\mathcal{T}_P$ and $\mathcal{T}_H$ may have diameters about 1\,mm.
  \item The tissue parenchyma including lower hierarchies of the ``inlet'' and
  ``outlet'' trees. The flow in this part of the vascular system is governed by
  the multi-compartment Darcy flow model in a porous medium. At any point of the
  liver occupying domain $\Omega \subset \mathbb{R}^3$, several (at least two)
  different pressure values are defined which are associated with different
  compartments, see Section~\ref{sec-0D}.  There are point sources and sinks
  defined in $\Omega$, where the 1D trees are connected to the parenchyma model.
\end{enumerate}

\subsection{Structure of compartments}

In this section we consider the lower hierarchies of the vascular trees.  We
consider a part of tissue (or the whole organ) occupying the domain $\Omega
\subset \mathbb{R}^3$.  The compartments are established as continuum
representations of the connected vascular network (or tree, as the special case)
restricted by a given range of characteristic vessel cross-sections. Denoting by
$N$ the total number of considered hierarchies, such a range is indicated by the
hierarchy index $j = 0,1,\dots,N$, whereby hierarchies $j = 0$ correspond to the
precapillary vessels of the lobular structure. Denoting by $a_j$ the typical (or
mean) vessel cross section in the given hierarchy, it holds that $i < j$
whenever $a_i < a_j$.  Each compartment is associated with the group; typically
there are two groups labeled by index $g = P, H$, corresponding to the portal
and hepatic vein systems (should also the hepatic artery be included in the
model, another group, say $g = A$ would be included). We shall identify
compartments by a simple index $k = \mathcal{C}(g,j)$ given by the group $g$ and
the hierarchy $j$, whereby $\mathcal{C}$ denotes the index set of all
compartments. Any compartment is represented by a subdomain $\Omega_k \subset
\Omega$ and by the hydraulic permeability tensor $\Kb^i$ which can be introduced
using an averaging procedure once the underlying tree segment (denoted by
$\mathcal{S}$) is known.  This point will be discussed below in
Section~\ref{sec-simulation}. For a given compartment $k$ belonging to a group
$g = \mathcal{G}(k)$, let $j = \mathcal{H}(k)$ is the hierarchy index.  The
fluid between two distinct compartments $i$ and $k$ can be exchanged only in the
overlap domain $\Omega_{ij} = \Omega_i \cap \Omega_k \not = \emptyset$, and
only, if either a) they belong to the same group $g = \mathcal{G}(k) =
\mathcal{G}(i)$ and their hierarchies are closed, $|\mathcal{H}(i) -
\mathcal{H}(k)| = 1$, or b) if $\mathcal{H}(i) = \mathcal{H}(k) = 0$ and
$\mathcal{G}(k) \not = \mathcal{G}(i)$. More general rules of flow between
compartments are possible, but would not be physiological, or would describe
special anomalies in the vasculature.

\paragraph{Groups, hierarchies and volume fractions}

The concept of volume fractions is a very natural basis to relate the model
parameters with structural features. Vessels of a given group $g$ and associated
with hierarchy $h$ are distributed at point $x \in \Omega$ with volume fractions
$\varphi_h^{g}(x)$. In the context of the present model, by $\phi_i(x)$ we
describe the local volume fraction of blood vessels comprised in compartment
$i$. Since $i = \mathcal{C}(g,h)$, obviously $\varphi_h^{g}\equiv \phi_i$.
Denoting by $\varphi_m$ is the volume fraction of all parts of the parenchyma
which are not blood vessels and, thus, not being included in any compartment,
the partition of unity expressed in terms of volume fractions of all components
reads in two possible ways:
\begin{align}\label{eq-ctp-vf}
  \sum_{h=0,\dots,N}\sum_{g = P,H}\varphi_h^{g}(x)
   + \varphi_m(x) & = 1\;,\quad
  \mbox{ or }\\\quad  \sum_i \phi_i(x) + \varphi_m(x) &= 1\;,
\end{align}
at any point $x \in \Omega$. Note that all volume fractions must be
non-negative.

\subsection{Multicompartment model of Darcy flow}\label{sec-0D}

We approximate the flow in lower hierarchies of the arterial and venous trees,
including  perfusion at the level of tissue parenchyma, using a macroscopic
model describing parallel flows in $\bar i$ defined compartments. The model is
based on some more fundamental principles related to homogenization, see Remark
in Section~\ref{sec-intro}, but can simply be justified by the phenomenological
approach. Similar models have been used in other works
\citep{Showalter2004,Cimrman2007,Michler2013}.

The model involves pressures $\{p^i\}$,  associated with each compartment $i =
1,\dots,\bar i$ . Any $i$-th compartment occupying domain $\Omega_i$ can be
saturated from an external source (or drained by a sink); in general one may
define the local source/sink flux $f^i$ which expresses an amount of fluid
supplied, or drained out from the compartment $i$ to the upper perfusion tree
branches, $\mathcal{T}_P$ and $\mathcal{T}_H$. Alternatively, the pressure can
be prescribed in a given subdomain $\Sigma_i \subset \Omega_i$ which may
represent junctions with upper hierarchies of the perfusion system treated by
the 1D flow model on branching networks, see Section~\ref{sec-Bf}. In practice,
$\Sigma_i$ can be formed by a number of ``small'' balls $\Sigma_i^k$, thus
$|\Sigma_i^k|<<|\Omega_i|$, labeled by index $k$ associated with the $k$-th
junction. The mass conservation for compartment $i$ is expressed by the
following equations (to be satisfied in $\Omega_i\setminus \Sigma_i$):
\begin{align}
  \nabla \cdot \wb^i + \sum_{j} \mathcal{J}_j^i & = f^i\;,
  \quad i = 1,\dots,\bar i \quad \mbox{ in } \Omega_i\setminus \Sigma_i\;,
  \label{eq-ctp-1a}
  \\
  \wb^i & = - \Kb^i\nabla p^i\;,
  \label{eq-ctp-1b}
  \\
  \mathcal{J}_j^i & = G_j^i(p^i - p^j)\;,
  \label{eq-ctp-1c}
\end{align}
where $\wb^i$ is the Darcy velocity, $\Kb^i = (K_{kl}^i)$ is the local
permeability tensor associated with vessels of the $i$-th compartment and
$G_j^i$ is the local perfusion coefficient coupling compartments $i,j$, so that
$\mathcal{J}_j^i$ describes the amount of fluid transported from $i$ to $j$
(drainage flux; obviously $\mathcal{J}_i^j = - \mathcal{J}_j^i$). As the
boundary conditions for \eq{eq-ctp-1a}, we consider the non-penetration
condition on the outer surface, and a prescribed pressure on $\pd\Sigma_i$,
\begin{align}
  \nb\cdot\wb^i & = - \nb\cdot\Kb^i\nabla p^i
    = 0\quad \mbox{ on }\pd \Omega_i\;,\label{eq-ctp-1bca}\\
  p^i & = \bar p^i\quad  \mbox{ on }\pd \Sigma_i,\; i
   = 1,\dots,\bar i\;.\label{eq-ctp-1bcb}
\end{align}

\begin{figure}
  \centering
  \includegraphics[width=0.65\linewidth]{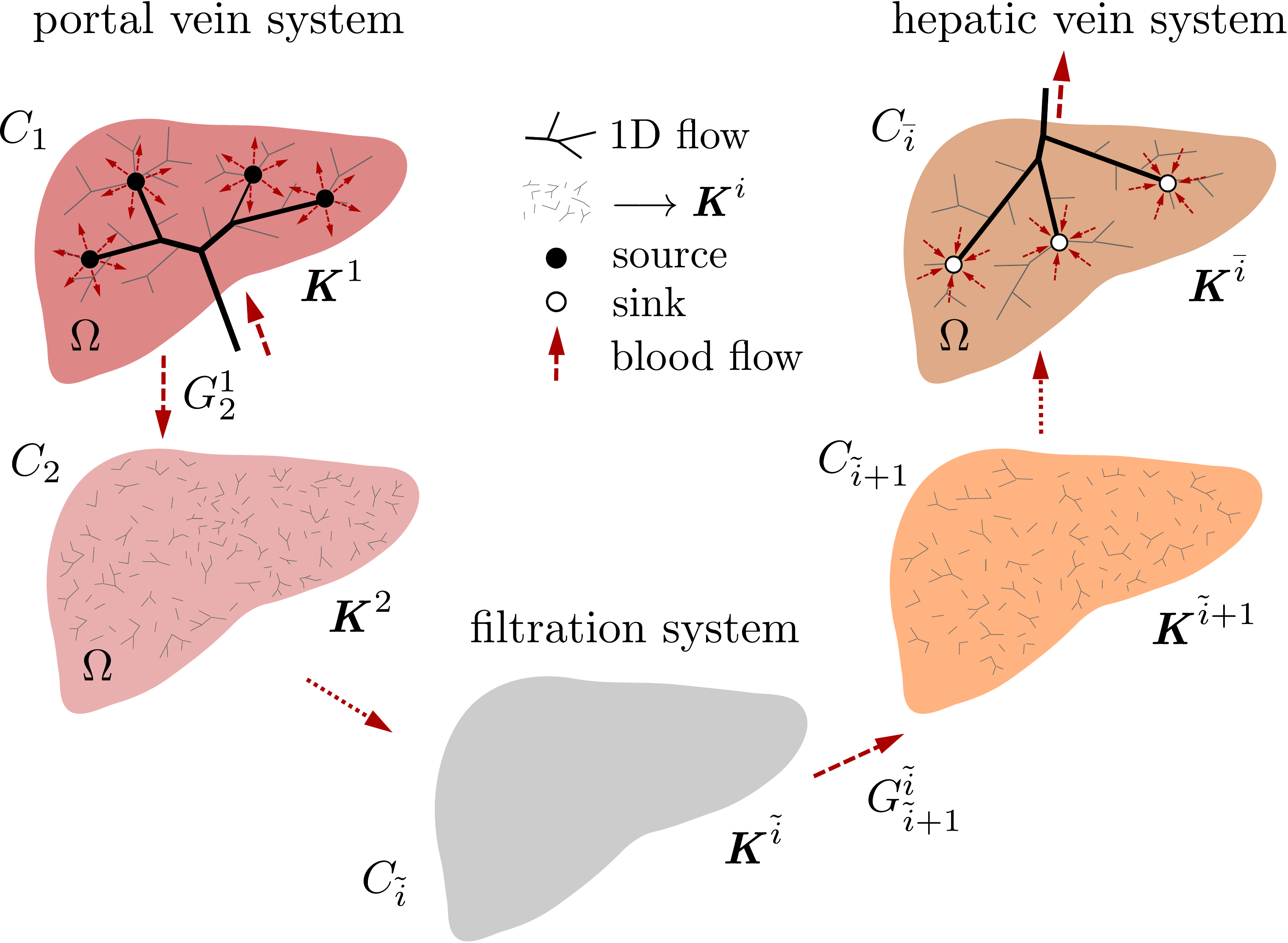}
  \caption{Schematic decomposition of the tissue perfusion system into
    compartments, note that the blood supply through the hepatic artery
    is not considered in our model
    }\label{fig-compartments}
\end{figure}

The numerical model is obtained by the finite-element discretization of the weak
formulation of problem \eq{eq-ctp-1a}-\eq{eq-ctp-1bcb}. We shall need the
following admissibility sets:
\begin{align}
  \mathcal{V}^i =
    \{q \in  H^1(\Omega_i)|\;q = \bar p_i \mbox{ on }\pd \Sigma_i\}\;,
    \label{eq-ctp-1asa}\\
  \mathcal{V}_0^i =
    \{q \in  H^1(\Omega_i)|\;q = 0 \mbox{ on }\pd \Sigma_i\}\;.
    \label{eq-ctp-1asb}
\end{align}
In our numerical tests we consider $\bar p_i$ being given by point values in
vertices of the finite element mesh. For some compartments, $\Sigma_i$ can
vanish, so that $\mathcal{V}^i = \mathcal{V}_0^i = H^1(\Omega_i)$.

\paragraph{The weak formulation of the problem} constituted by equations
\eq{eq-ctp-1a}-\eq{eq-ctp-1c} and the boundary conditions
\eq{eq-ctp-1bca}-\eq{eq-ctp-1bcb} is, as follows:
Find $\bmi{p} = (p^1,p^2,\dots,p^{\bar i})$ with $p^i \in \mathcal{V}^i$, such
that for all compartments $i = 1,\dots,\bar i$, 
\begin{multline}
  \label{eq-wf}
  \int_{\Omega_i\setminus \Sigma_i} \Kb^i \nabla p^i \cdot \nabla q^i
   + \int_{\Omega_i\setminus \Sigma_i} \sum_j G_{j}^i( p^i - p^j) q^i\\
   = \int_{\Omega_i\setminus \Sigma_i}  f^i q^i\;,
    \quad \forall q^i \in \mathcal{V}_0^i \;.
\end{multline}
The summation in the second term takes only over nonvanishing $G_{j}^i$.

All parameters $\Kb^i(x)$ and $G_{j}^i(x)$ depend on the local volume fractions
$\phi_i(x)$. In fact, for a given tree, see Fig.~\ref{fig-ph_trees}, both these
parameters can be computed point-wise in $\Omega$ using an averaging approach
reported in \cite{vankan-huyghe_1997,Michler2013} which is based on the
theoretical result describing hierarchical flows \citep{huyghe_campen_1995:ALL}.
Then the following properties can be guaranteed:
\begin{itemize}

  \item $\Kb^i(x)$ is positive semi-definite 2nd order tensor for $x \in
  \Omega$.

  \item Domain $\Omega_i = \{x \in \Omega\,,\,|\Kb^i(x)| > 0 \}$.

  \item $G_i^j(x)\geq 0$ for $|i-j| = 1$ in the overlap subdomains $\Omega_i
  \cap \Omega_j \not = \emptyset$.

\end{itemize}
Usually, when the averaging control volume $V_c$ is large enough {w.r.t.{~}} the
characteristic size of the vessel segments, $\Kb^i$ is positive definite,
however, if only one or two vessels are encountered in $V_c$, $\Kb^i$ is
singular. For the numerical treatment, a regularization can be considered,
{e.g.{~}} using modified permeabilities $\tilde \Kb^i(x) := \Kb^i(x) +
\varepsilon \bar K \bmi{I}$, where $\bmi{I} = (\delta_{ij})$, $\bar K =
\|\Kb^i\|_{\Omega_i}$, but can be any relevant permeability value in domain
$\Omega_i$, and $\varepsilon > 0$ is a small number.

\paragraph{Well posedness of the perfusion problem.}
Assume nonvanishing  Lipschitz open bounded domains $\Omega^i$, such that for
each $i = 1,\dots,\bar i-1$ there exists a nonempty overlap domain
$\Omega_{i,i+1} = \Omega_i\cap \Omega_{i+1} \not = \emptyset$ and $G_{i+1}^{i} >
0$ in $\Omega_{i,i+1}$. If at least for one compartment $j \in \{ 1,\dots,\bar
i\}$ the Dirichlet boundary $\pd \Sigma_j$ is nonempty, then the regularized
problem \eq{eq-wf}, where $\Kb^i$ is replaced by $\tilde \Kb^i$, yields a unique
solution $\tilde {\bmi{p}}$.

\subsection{Flow on upper-level vascular trees}\label{sec-Bf}

As announced above, the upper-level vascular trees, the arterial, or the venous
one, should be treated by taking into account the specific geometry of the
vessels. In order to obtain an efficient numerical model it has been suggested
to consider a simplified flow model based on the Bernoulli equation. Thus,
instead of the full CFD analysis of flow in complicated 3D vessel geometries,
the perfusion tree is replaced by a system of line segments characterized by the
length, the cross-section and a loss parameter which is related to the specific
geometry of the vessel segment. To respect the tortuosity of the vascular
network, the lengths associated with the line segments should correspond to the
actual vessel length.

\begin{figure}
  \centering
  \includegraphics[width=0.48\linewidth]{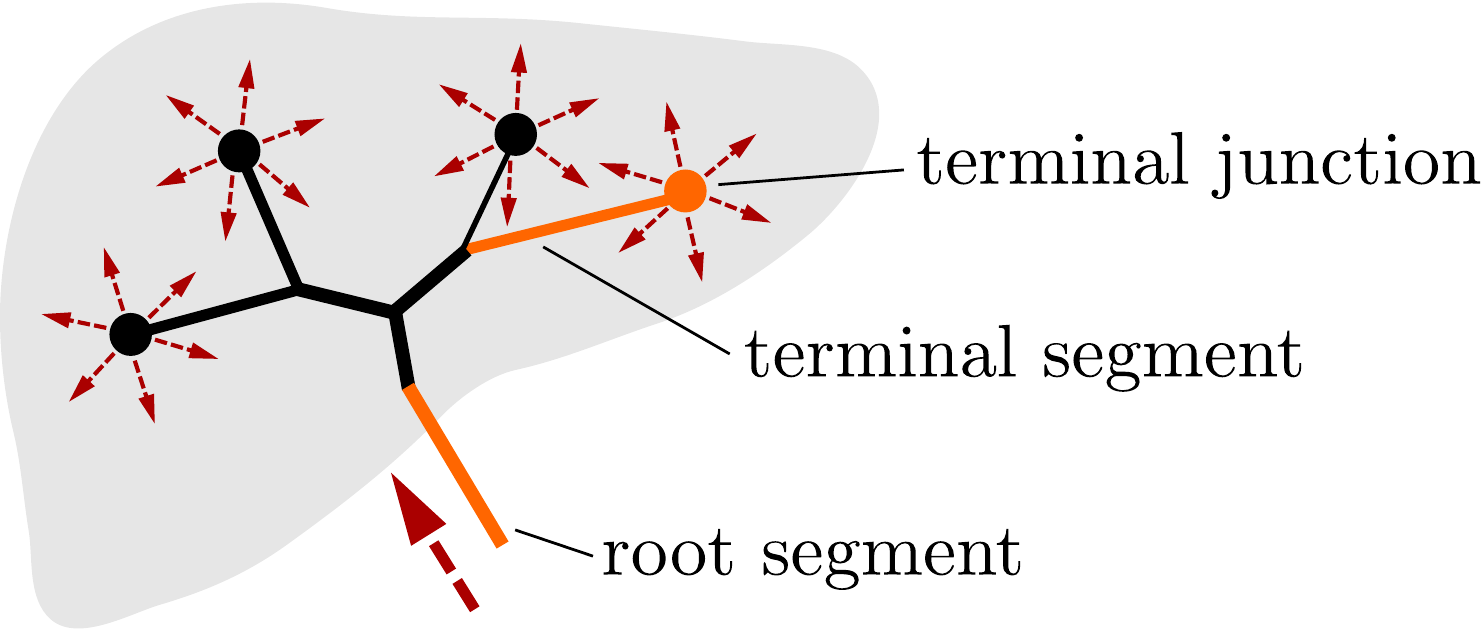}\\
  \caption{Structure of the 1D vascular tree}
  \label{fig-tree_1D}
\end{figure}

We consider a vascular tree $\mathcal{T}(\{\mathcal{B}^j\}_j,\{\ell_e\}_e)$
constituted by vessel segments $\ell_e$ and junctions $\mathcal{B}^j$ labeled
by $j$, see Fig.~\ref{fig-tree_1D}. Any $\mathcal{B}^j = (X^j,J^j)$ is defined
by its spatial position $X^j \in \mathbb{R}^3$ and by the connectivity set $J^j$
containing indices of segments connected at the junction. Although it is assumed
that $\mathcal{T}$ may form a graph with loops, in our study we consider only
branching trees, so that $\mathcal{B}^j$ is typically a bifurcation. The root
and terminal junctions are just one-element sets. The number of all terminal
junctions is denoted by $\hat n$. By $J^0$ we denote the root junction, whereas
the terminal branches end by junctions $\hat J^k$ through which they are
connected with the continuum model described above in Section~\ref{sec-0D}.

The simplest possible model of flow on $\mathcal{T}$ is based on the Bernoulli
equation. The flow in a tree $\mathcal{T}$ is described by nodal pressures
$\{p_j\}$ and segment velocities $\{w_e\}$. In order to take into account the
dissipation of the viscous flow, the pressure loss terms must be introduced by
virtue of the Reynolds number associated with the flow on one segment. For any
line segment $\ell_e$ featured by a ``true'' vessel length $L_e$ and its
diameter $D_e$ which is assumed to be constant along the whole segment, the
pressure drop related to the node pressures at the vessel end-points is related
to the velocity $w_e$ by the following relationship
\begin{equation*}
  \Delta_e p = p_i - p_j =  \frac{1}{2}\rho
   \lambda_e w_e^2\;, \quad e \in J^i \cap J^j\;,
\end{equation*}
where $\rho$ is the blood density and $\lambda_e =
\frac{64}{\textsf{Re}(w_e)}\frac{L_e}{D_e}$ is derived using the Poiseuille flow
model; note that the Reynolds number depends on the velocity $w_e$.

The flow model for the tree $\mathcal{T}$ is assembled using the mass
conservation (the continuity equation) expressed for each junction $j$,
\begin{equation}\label{eq-Be1}
  \sum_{e \in J^j}   A_e w_e = 0\;,\quad j = 1,\dots\;,
\end{equation}
where $A_e = \pi D_e^2/4$ is the cross-section, and by the modified Bernoulli
equation written for all branches of any bifurcations $j$,
\begin{equation}\label{eq-Be2}
\frac{1}{2}\rho w_{e_i}^2 + p_i = \frac{1}{2}\rho w_{e_j}^2 + p_j
  + \frac{1}{2}\rho \lambda_{e_j} w_{e_j}^2\;,
\end{equation}
where $e_j$  is the segment index of vessel $\ell_{e_j}$ connecting nodes $i$
and $j$, \ie $e_j \in J^i \cap J^j$, whereas ${e_i} \in J^i$ is the source
vessel of node $i$. The final system of equations involving all bifurcations is
solved by  the Newton method.

\subsection{Coupling the 1D and the multicompartment Darcy flow models}
\label{sec-alg}

The 1D model is coupled with the  model introduced in Section \ref{sec-0D}
through the terminal junctions which specify the sources and sinks $f^i$ for all
the considered compartments of the upper-most hierarchy, \ie for $i=1$ or $i =
\bar{i}$. For the $i$-th compartment saturated by the tree $\mathcal{T}$ we
define
  \begin{align}
    f^i(x) & = \sum_{k = 1}^{\hat n}\delta(x - \hat x^k) A_k w_k\;,
      \label{eq-ssfa}\\
      p^i(x) & = p_k\;,\quad x \in \Sigma_i^k\;,
      \label{eq-ssfb}
  \end{align}
where $\delta(x - \hat x^k)$ is the Dirac distribution centered at point $\hat
x^k \in \Omega$ which is associated with the terminal junction $k$  of the 1D
model. The pressures at these junctions represented by $\Sigma_i^k$ are coupled
with the pressure fields in parenchyma, as expressed by condition \eq{eq-ssfb}.
In practice, we use an approximation of $\delta(x - \hat x^k)$ which is based
upon the specific finite element discretization, and \eq{eq-ssfb} is replaced by
$p^i(\hat x^k) = p_k$ so that only nodal value of the pressure is shared.

We conclude by a simple iterative algorithm used to compute the perfusion
pressure and velocity fields in a steady state which is reached by increments
for a pseudo-time step. Since there are two trees $\mathcal{T}$, as illustrated
in Fig.~\ref{fig-sources_sinks}, we shall label the corresponding solutions by
indices $P$ and $H$, denoting quantities associated with the portal and hepatic
veins. For given values $\bar w_0^P$ and $\bar p_0^H$, \ie the velocity in the
inlet portal vein and the pressure in the (outlet) hepatic vein, the computation
proceeds by repeating the following algorithm:
\begin{enumerate}
  \item Set all interface velocities and pressures to zero, namely $\{p_k^P\} =
  0$ and $\{w_k^H\} = 0$. Set $i = 0$ and $\tau = 1/N$, for a given $N \in
  \mathbb{N}$, the number of pseudo-time steps.

  \item For the new iteration $i:=i+1$, update $w_0^P:= \min\{i\tau,1\} \bar
  w_0^P$ and $p_0^H:= \min\{i\tau,1\} \bar p_0^H$.

  \item Solve \eq{eq-Be1}-\eq{eq-Be2} on trees $\mathcal{T}_P$ and
  $\mathcal{T}_H$, so that\\
   $\left( w_0^P,\{p_k^P\}\right) \mapsto \left(
  p_0^P,\{w_k^P\}\right)$, for the artery, and\\
   $\left( p_0^H,\{w_k^H\}\right)
  \mapsto \left( w_0^H,\{p_k^H\}\right)$, for the vein.

  \item Solve \eq{eq-ctp-1a}-\eq{eq-ctp-1bcb} in $\Omega$, so that\\
  $\left(\{w_k^P\},\{p_k^H\}\right) \mapsto \left(\{p_k^P\},\{w_k^H\}\right)$
  and $(p^i(x),\wb^i(x))$ is computed for $x \in \Omega$.

  \item Use the conditions \eq{eq-ssfa}--\eq{eq-ssfb} to update the interface
  variables for the next iteration $i+1$.

  \item Go to step 2, unless a steady state is reached.
\end{enumerate}


\section{Contrast fluid transport}\label{sec-CT}

In this section we explain how to simulate the dynamic perfusion tests which are
used as a principal method to assess blood flow in highly perfused organs,
{e.g.{~}} in the liver, or brain. It is based on {the} computed tomography
{scans which provide information on} \emph{the tissue density}. This quantity is
proportional to the local concentration of the contrast fluid (the tracer)
which is dissolved
in the blood. Its relative content is expressed by the saturation $S$. This
quantity is defined for each of the individual compartments of the parenchyma
treated by the Darcy flow model and also for all branches of the upper level
vascular trees, as explained in Section~\ref{sec-Bf}.

Transport equations for resolving the saturations in all the compartments can be
derived using the pre-computed perfusion velocities, the mass conservation law,
and taking into account fluid exchange between the compartments. Thus, we obtain
a system of hyperbolic equations for resolving the saturations. Then the tissue
contrast is defined locally as the weighted sum of all the saturations; the
weights are given by the volume fractions.


\subsection{Tracer distribution in compartments}

The tracer saturation associated with the $i$-th compartment is denoted by
$S^i$; its values are restricted by $S^i \in [0,1]$. This restriction must be
guaranteed by the transport equations arising from the conservation law whenever
the feasible initial and boundary conditions satisfy this constraint. Then we
can introduce the tracer partial concentration, $c^i = \phi^i S^i$ (no
summation) for each compartment $i$. This quantity is equivalent to the
tissue density retrieved from CT scans.

The total apparent concentration corresponding to the grey levels is then given as
\begin{equation}\label{eq-ctp-2}
  C = \sum_i c^i = \sum_i \phi^i S^i\;,
\end{equation}
where the summation is taken over all locally overlapping compartments.

The local conservation in domain $\Omega_i \subset \Omega$ for the $i$-th
compartment is expressed, as follows:
\begin{multline}\label{eq-ctp-3}
  \int_{\Omega_i} \phi^i \frac{\partial S^i}{\partial t}
  + \int_{\pd \Omega_i} \wb^i \cdot \nb S^i\, \mathrm{d} \Gamma
  + \sum_j \int_{\Omega_i} Z_j^i(S)\mathcal{J}_j^i\\
  = \int_{\Omega_i} S_{\rm in} f_+^i + \int_{\Omega_i} S^if_-^i \;,
\end{multline}
where $S_{\rm in}$ is the external source saturation, $f_+^i>0$, the positive
part of $f^i$ is the in-flow (while the negative part $f_-^i$ is the out-flow) and the $Z_j^i(S)$ is
the nonlinear operator defined, as follows (note $\mathcal{J}_j^i$ is given by
\eq{eq-ctp-1c}):
\begin{equation}\label{eq-ctp-4}
  Z_j^i(S) = \left \{
  \begin{array}{ll}
    S^i & \mbox{ if } \mathcal{J}_j^i>0\;,\\
    S^j & \mbox{ if } \mathcal{J}_j^i\leq 0\;.
  \end{array}\right .
\end{equation}
From \eq{eq-ctp-3} we deduce the following problem: given $\{\wb^i\}$,
$\{p^i\}$ and initial conditions $\{S^i(t=0,x)\} = \{S_0^i(x)\}$ defined
in $\Omega_i$, find $\{S^i(t,x)\}$ such that
\begin{multline}\label{eq-ctp-5}
  \phi^i \frac{\partial S^i}{\partial t} + \nabla \cdot (S^i \wb^i)
   + \sum_j Z_j^i(S)\mathcal{J}_j^i =
  S_{\rm in} f_+^i + S^if_-^i\\
  \quad x \in \Omega_i, \quad t > 0\;, \quad i = 1,\dots \bar i\;,
\end{multline}
\begin{equation*}
  S^i \mbox{ given on }  \pd_{i-} \Omega_i(\wb^i)\;,
\end{equation*}
where $\pd_{i-} \Omega_i(\wb^i) = \{x \in \pd \Omega_i|\; \wb^i\cdot \nb < 0\}$
is the in-flow boundary of $\Omega_i$. However, $\pd_{i-} \Omega_i(\wb^i) =
\emptyset$ in our problem due to the boundary condition \eq{eq-ctp-1bca}.

Instead of the switch $Z$ we may introduce corresponding index sets:
\begin{equation}\label{eq-ctp-6}
  \mathcal{I}_+^i = \{j\not = i|\; \mathcal{J}_j^i > 0\}\;,\quad
  \mathcal{I}_-^i = \{j\not = i|\; \mathcal{J}_j^i \leq 0\}\;.
\end{equation}
Further, by introducing the ``true mean velocities'' $\vb^i =
(\phi^i)^{-1}\wb^i$, we can rewrite \eq{eq-ctp-5}, as follows:
\begin{equation}\label{eq-ctp-6a}
  \phi \frac{D_{\vb^i} S^i}{D t} + S^i \nabla\cdot \wb^i +
  \sum_{j \in \mathcal{I}_-^i} S^j \mathcal{J}_j^i
   +\sum_{j \in \mathcal{I}_+^i} S^i \mathcal{J}_j^i
  = S_{\rm in} f_+^i + S^i f_-^i \;,
\end{equation}
where $\frac{D_{\vb^i} S^i}{D t}$ is the material derivative {w.r.t.{~}}
$\vb^i$.

\subsection{Transport on branching network}

The upper hierarchies of the perfusion trees $\mathcal{T}_P$ and $\mathcal{T}_H$
are represented by branching networks consisting of \emph{vessel segments} and
\emph{junctions}. For such structures we can derive the transport (advection)
equations. By $x$ we refer to the axial coordinate along the oriented line
segment $\ell = ]x_0,x_1[$, $x_0,x_1 \in \mathbb{R}$ with the end-points
$x_0,x_1$, while by $X \in \mathbb{R}^3$ we mean the spatial positions
associated with $x$. We consider a velocity $w(x)$ and cross-section $A(x)$
given at any $x \in \ell$, which satisfy the mass conservation (by $Q$ we denote
the flux in the segment)
\begin{equation}\label{eq-ctp-10}
  \pd_x (w A) = \pd_x Q = 0\;, \quad x \in \ell.
\end{equation}
The positiveness of the convection velocity $w$ is established in the context of
the orientation of the vessel segment.

It is now easy to derive the following equation for transport of the tracer,
where $S(x,t)$ is the local instantaneous saturation; possible forms of the same
equation are:
\begin{equation}\label{eq-ctp-11}
  \pd_tS(x,t) + w(x)\pd_x S(x,t) = 0\;,
  \quad x \in \ell.
\end{equation}
At the line segment ends we consider the boundary conditions:
\begin{align}
  S(x_0,t) &= S_0(t)\quad \mbox{ given for } Q >0\;, \label{eq-ctp-12a}\\
  S(x_1,t) &= S_1(t)\quad \mbox{ given for } Q \leq 0\;\label{eq-ctp-12b}.
\end{align}

\paragraph{Transition times.}
We consider given saturations $S_0(t)$ and $S_1(t)$ at the end-points of vessel
segment $\ell_e$, see \eq{eq-ctp-12a}--\eq{eq-ctp-12b}.  From  \eq{eq-ctp-11}
one can obtain the following equation:
\begin{equation}\label{eq-ctp-13}
  S(x_1, t_0 + T_e) = S(x_0,t_0)\;,
\end{equation}
where $T_e = \int_{x_0}^{x_1} (w(x))^{-1}\, \mathrm{d} x$ is the transition time
of the transport between the endpoints.

\paragraph{Mixing and transport through junctions.}
At any junction, a unique saturation $\tilde S^j$ is computed using an obvious
conservation law.  The mean junction saturation $\tilde S^j$ satisfies
\begin{equation}\label{eq-ctp-14}
  \sum_{e \in J_+^j} S_e A_e v_e^j
   + \tilde S^j \sum_{e \in J_-^j} A_e v_e^j = 0\;,
\end{equation}
where the two index sets $J_-^j$ and $J_+^j$ are defined according to the flow
orientation in the vessel segments passing through the junction $j$, as follows:
\begin{align}
  v_e^j &= +w_e \qquad \mbox{ if $e$ ends at junction } j\;,\nonumber\\
  v_e^j &= -w_e \qquad \mbox{ if $e$ begins at junction } j\;,\nonumber\\
  J_+^j &=  \{ e \in J^j|\, v_e \geq 0\}\;,\nonumber\\
  J_-^j &=  \{ e \in J^j|\, v_e < 0\}\;,\nonumber
\end{align}
so that velocities $w_e$ in the vessel segment $\ell_e$ define $v_e$ depending
on the oriented network topology. We can call $J_+^j$ the index set of sources
and $J_-^j$ the index set of sinks. It is worth noting that for the perfusion
tree $\mathcal{T}_P$ (portal vein tree) with binary bifurcation at any junction,
the set $J_+^j  = e$ contains just one index of the saturating vessel, while
$J_-^j$ contains the two children vessels, say $e_1,e_2$. For the hepatic vein
tree $\mathcal{T}_H$ with the analogous property, the role of $J_-^j = e$ and
$J_+^j = e_1,e_2$ is exchanged.

\paragraph{Assembling equations of the transport on the network.}
Due to \eq{eq-ctp-13} and knowledge of the transition times, the state of the
transport is described by the junction saturations $\{\tilde S^j(t)\}_j$. The
resulting system of equations governing the nodal saturations takes the
following form:
\begin{equation}\label{eq-ctp-15}
  \tilde S^j(t) \sum_{e \in J_-^j} A_e v_e^j +
  \sum_{e \in J_+^j} A_e v_e^j \tilde S^{k_e}(t - T_e) = 0\;.
\end{equation}


The junction equations \eq{eq-ctp-15} can be evaluated for discretized time
interval, \ie for $t \in \{t_n\}_n$ where $t_n = t_0 + n \Delta t$. Obviously,
for a given $T_e$, the saturation at $t_n - T_e \in [t_p,t_{p+1}]$ is
approximated using the average of values at $t_p$ and $t_{p+1}$ .

\section{Numerical simulation of liver perfusion}\label{sec-simulation}

The model introduced in the preceding sections has been implemented in our
non-commercial codes. In this section,  we illustrate the model response using
numerical examples with real geometry of the human liver, but with the perfusion
trees of the portal and hepatic veins generated using the CCO method
\citep{gco}. Before presenting particular results which serve as a proof
concept, we explain a  flowchart of the currently developed computational
modeling tool.

CT scans of human liver processed by the {\it LISA (LIver Surgery Analyser)}
software \citep{LISA} are the starting point to the numerical simulation of
liver perfusion. The {\it LISA} code includes the semi-automated segmentation
method based on the graph-cut algorithm and returns the geometrical model of
liver parenchyma that can be transformed into a volumetric finite element mesh.
The code is also used to identify the position and diameter of the portal and
hepatic veins at the point of entry into the liver parenchyma; this data is
necessary to generate two artificial vascular trees associated with the portal
and hepatic veins, which in part are employed directly in the simulation of the
upper-level flow using the 1D flow model described in Section~\ref{sec-Bf}, but
also to establish permeabilities $\Kb^i$ and perfusion coefficients $G_i^j$
appearing in the multicompartment Darcy flow model. The hepatic artery is not
considered in our liver model because of its small contribution to the overall
blood perfusion through the liver tissue. A constrained constructive
optimization approach is used for the creation of vascular structures reflecting
main physiological principles, see \cite{gco}, and the geometry of a particular
liver.

\begin{figure}
  \centering
  \includegraphics[width=0.95\linewidth]{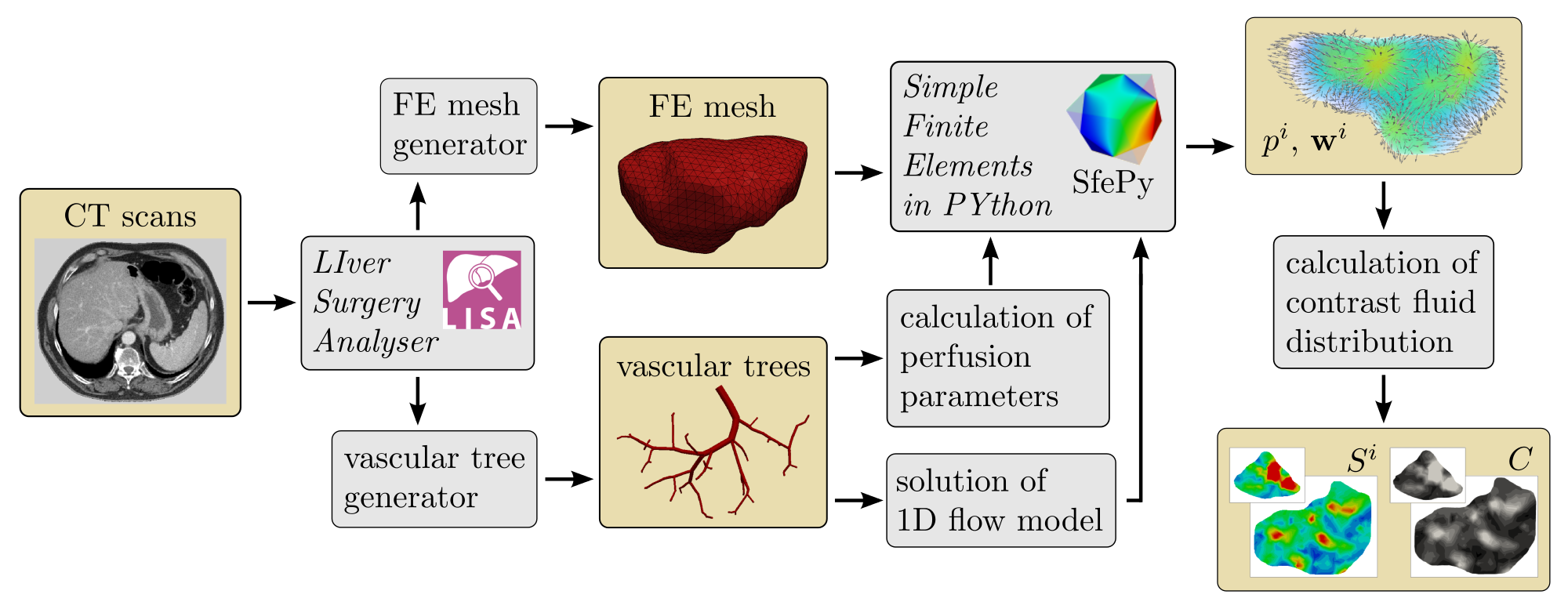}
  \caption{Flowchart of the liver perfusion simulation process}
  \label{fig-comp_workflow}
\end{figure}

In accordance with the modeling principles described above, the calculation of
blood perfusion and contrast fluid distribution in the liver can be divided into
two stages. The first one includes the  solution of 1D flow model
\eq{eq-Be1}-\eq{eq-Be2} coupled with the multicompartment Darcy model
\eq{eq-wf}. 
The resulting velocities associated with vessel segments of the 1D flow model
and velocity fields $\wb^i$ in compartments $i$ are then employed in the second
stage in which the compartmental saturations $S^i$ yielding  the total
concentration $C$ are computed. The 1D flow model is implemented in the {\it
Python} language, the multicompartment Darcy flow model is discretized using the
FE method in software {\it SfePy -- Simple Finite Elements in Python}, see
\cite{cimrman_2014:sfepy}. The model describing the tracer transport is
implemented using the finite volume method in the {\it Matlab} system. The
overview of the simulation process is illustrated in
Fig.~\ref{fig-comp_workflow}.

\begin{figure}
  \centering
  \includegraphics[width=0.85\linewidth]{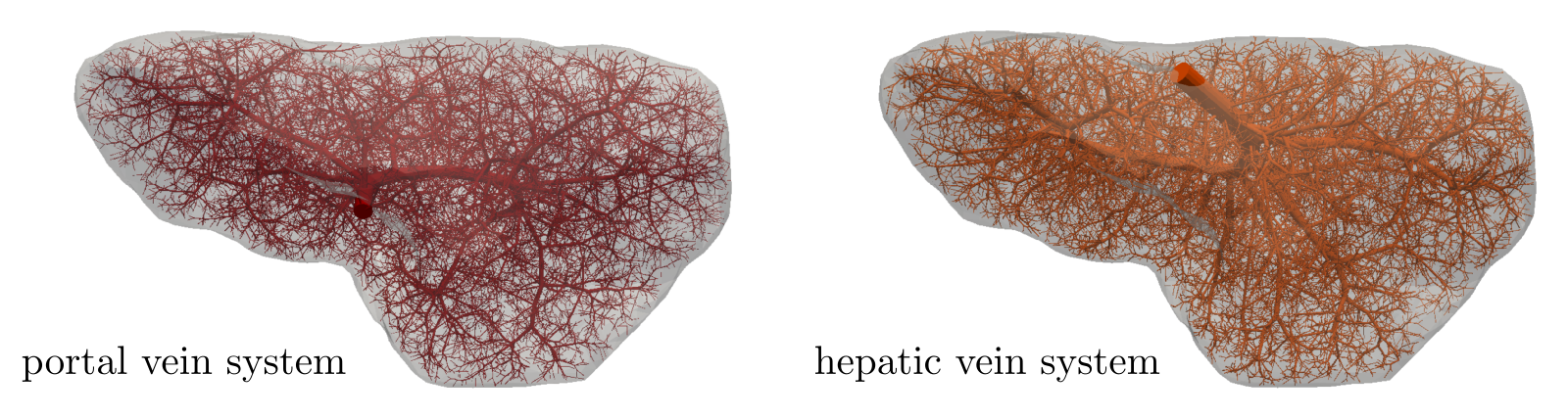}
  \caption{Artificially generated vascular trees
    (3D visualization of tree structure)
    representing the portal {\it(top)} and hepatic {\it(bottom)}
    vein systems}
  \label{fig-ph_trees}
\end{figure}

\begin{figure}
  \centering
  \includegraphics[width=0.60\linewidth]{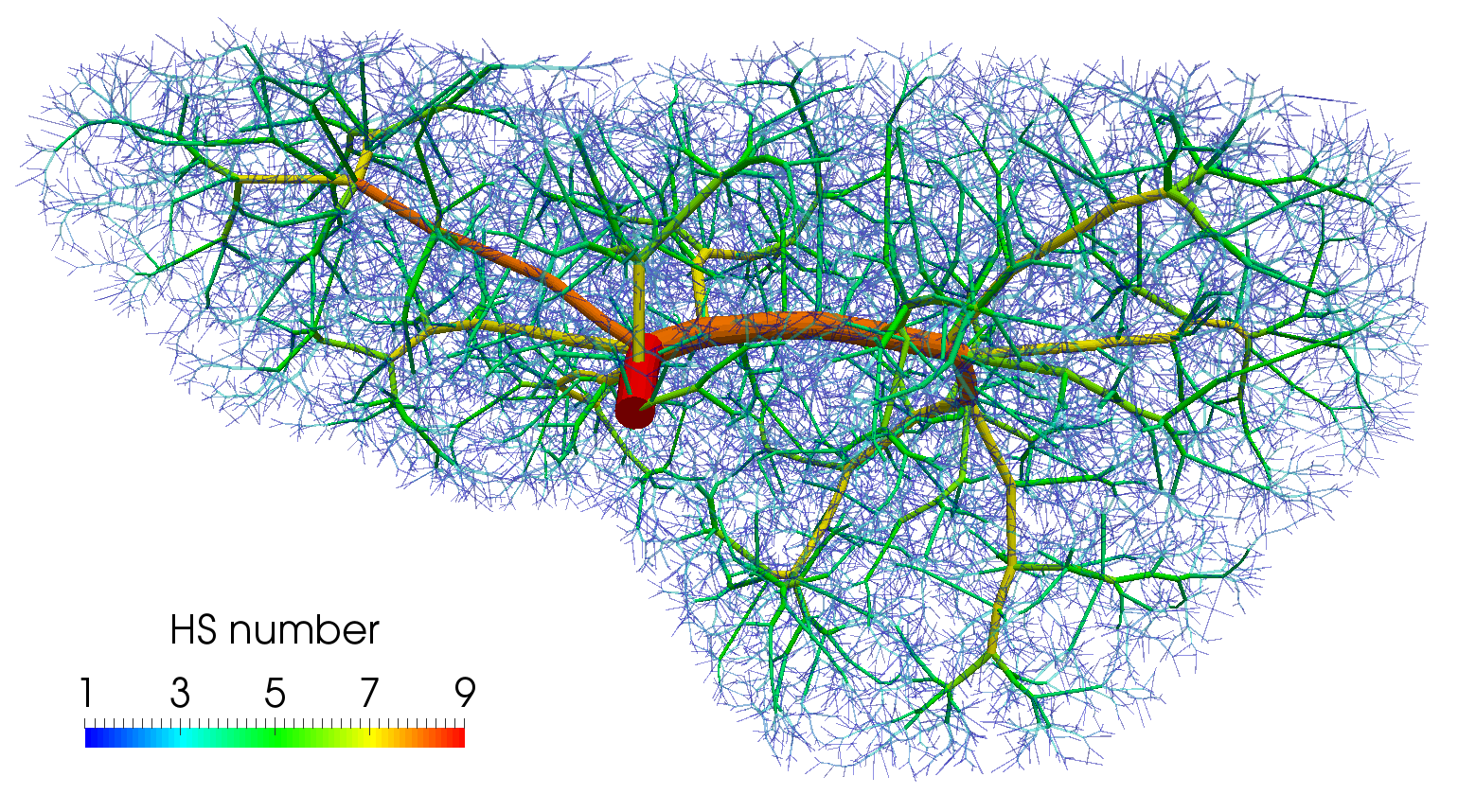}
  \caption{The artificial portal vein tree and its branching complexity
    measured by the Horton-Strahler number}
  \label{fig-hs_portal_tree}
\end{figure}

To demonstrate features of the hierarchical model described in this paper, a
three compartment model is considered. The first compartment represents the
vascular network of the portal vein consisting of veins with diameter in the
range of about $10^{-3}- 10^{-4}$\,m. The same diameter range is chosen also for
the third compartment which stands for the system of the hepatic vein.
Compartment 2 involves small veins of diameters smaller than $10^{-4}$\,m
including the hepatic capillaries of the lobular sinusoids, and it may therefore
be perceived as a filtration part of the perfusion model. The generated
artificial vascular trees representing the vascular networks in the liver are
depicted in Fig.~\ref{fig-ph_trees}, their properties are summarized in
Tab.~\ref{tab-ph_trees}. The higher hierarchy trees, where the blood flow is
approximated by the 1D model, see Fig.~\ref{fig-multiperf}, are obtained by
taking the vessel segments with the Horton--Strahler (HS) number \citep{gco}
greater or equal to 6. The branching complexity of the generated portal vascular
tree characterized by the HS number is illustrated in
Fig.~\ref{fig-hs_portal_tree}.

\begin{table}
    \begin{tabular}{c|c|c|c|c}
      vascular
      & number of all & number of & diameter of & diameter of\\
      tree
      & vessel segments & terminal segments & the root segment &terminal segments\\
      \hline
      portal & 36\,739 & 19\,777 & 8.4\,mm & $\approx 10^{-4}$\,mm \\
      \hline
      hepatic & 37\,133 & 19\,769 & 5.8\,mm & $\approx 1.4 \times 10^{-4}$\,mm \\
    \end{tabular}
    \caption{Properties of artificial vascular trees}\label{tab-ph_trees}
\end{table}

To determine parameters $\Kb^1$, $\Kb^3$ and $G_1^2$, $G_2^3$, as well as the
porosities $\phi^1$, $\phi^3$ in each finite element, a volume averaging is
performed over the corresponding vascular tree for a given range of the vessel
diameters. The anisotropic permeabilities $\Kb^1$, $\Kb^3$ of compartments 1~and
3 are evaluated using the expression derived by Huge and van Campen
\citep{huyghe_campen_1995:ALL}, whereas the coupling perfusion coefficients
$G_1^2$, $G_2^3$ are calculated with the help of the theory presented in
\cite{Michler2013}. Values of the model parameters at two selected points are
listed in Tab.~\ref{tab-perf_params}. The isotropic permeability in
compartment~2 is taken as $K^2 = 2\times 10^{-14}$ and the uniform porosity as
$\phi^2 = 0.15$, both values are chosen with respect to the parameters
identified in~\cite{Debbaut2012}. The domain of the liver is discretized by a
finite element mesh consisting of 11\,019 tetrahedrons and 2\,442 nodes. The
rather coarse mesh was chosen due to large computational cost associated with
the tracer transport calculation.

\begin{table}
    \begin{tabular}{c|c|c|c}
      point &
      permeability $\Kb$ [m$^2\cdot$(Pa$\cdot$s)$^{-1}$]
      & porosity $\phi$ [--]
      & coef. $G$ [(Pa$\cdot$s)$^{-1}$]\\
      \hline
      $\hat{\textrm A}$&
      $\Kb^1 =
        \begin{pmatrix}
          \begin{matrix}
             1.10 & -0.01 & 0.12\\
             -0.01 & 1.21 & 0.08\\
             0.12 & 0.08 & 1.11
          \end{matrix}
        \end{pmatrix} \times 10^{-9}$
        & $\phi^1 = 1.05 \times 10^{-3}$
        & $G_1^2 = 3.52 \times 10^{-5}$\\
      &
      $\Kb^3 =
        \begin{pmatrix}
          \begin{matrix}
             4.38 & -0.18 & -0.61 \\
             -0.18 & 3.84 & -0.03 \\
             -0.61 & -0.03 & 4.81
          \end{matrix}
        \end{pmatrix} \times 10^{-9}$
        & $\phi^3 = 1.76 \times 10^{-3}$
        & $G_2^3 = 4.27 \times 10^{-5}$\\
        \hline
        $\hat{\textrm B}$&
        $\Kb^1 =
          \begin{pmatrix}
            \begin{matrix}
              1.00 & -0.03 & 0.11 \\
              -0.03 & 1.05 & 0.08 \\
              0.11 & 0.08 & 0.95
            \end{matrix}
          \end{pmatrix} \times 10^{-9}$
          & $\phi^1 = 0.93 \times 10^{-3}$
          & $G_1^2 = 3.26 \times 10^{-5}$\\
        &
        $\Kb^3 =
          \begin{pmatrix}
            \begin{matrix}
              4.32 & -0.15 & -0.60 \\
              -0.15 & 3.55 & -0.02 \\
              -0.60 & -0.02 & 4.80
            \end{matrix}
          \end{pmatrix} \times 10^{-9}$
          & $\phi^3 = 1.64 \times 10^{-3}$
          & $G_2^3 = 4.20 \times 10^{-5}$\\
    \end{tabular}
    \caption{
      Perfusion parameters at two distinct points $\hat{\textrm A}$,
      $\hat{\textrm B}$  (see Fig.~\ref{fig-liver_sec_AB}) related to
      compartments 1 and 3 obtained on the basis of calculations carried out on
      the artificial vascular trees}\label{tab-perf_params}
\end{table}

\begin{figure}
  \centering
  \includegraphics[width=0.8\linewidth]{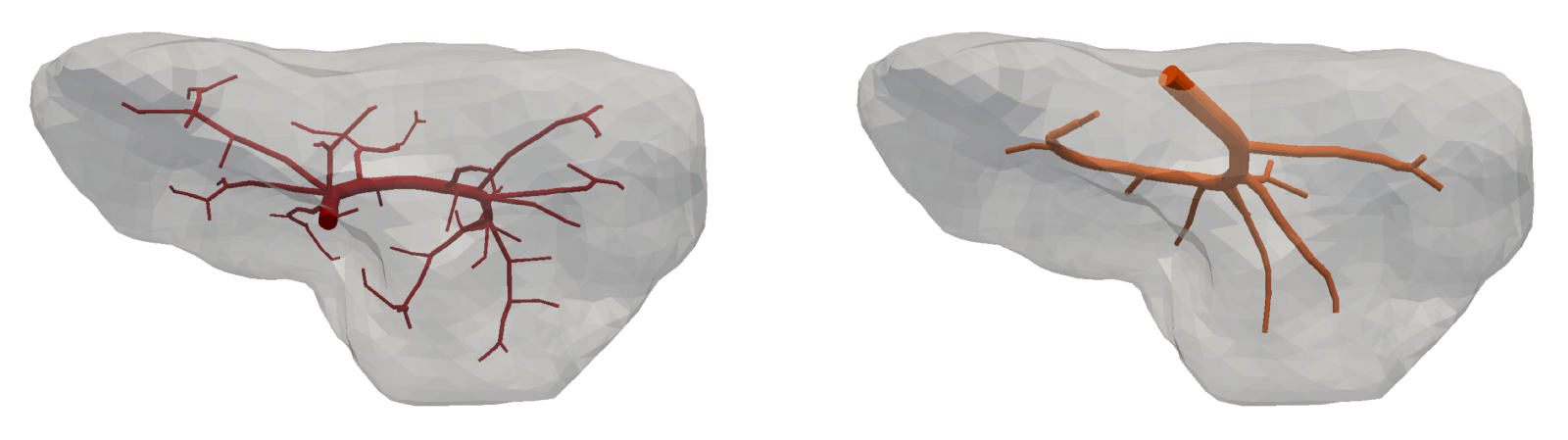}
  \caption{
    The upper-level vascular trees ({\it top} -- portal, {\it bottom} --
    hepatic) where the flow is approximated using the 1D model based on the
    extended Bernoulli equation. Their terminal points correspond either to the
    sources in compartment 1 or to the sinks in compartment 3}
  \label{fig-sources_sinks}
\end{figure}

\begin{figure}
  \centering
  \includegraphics[width=0.98\linewidth]{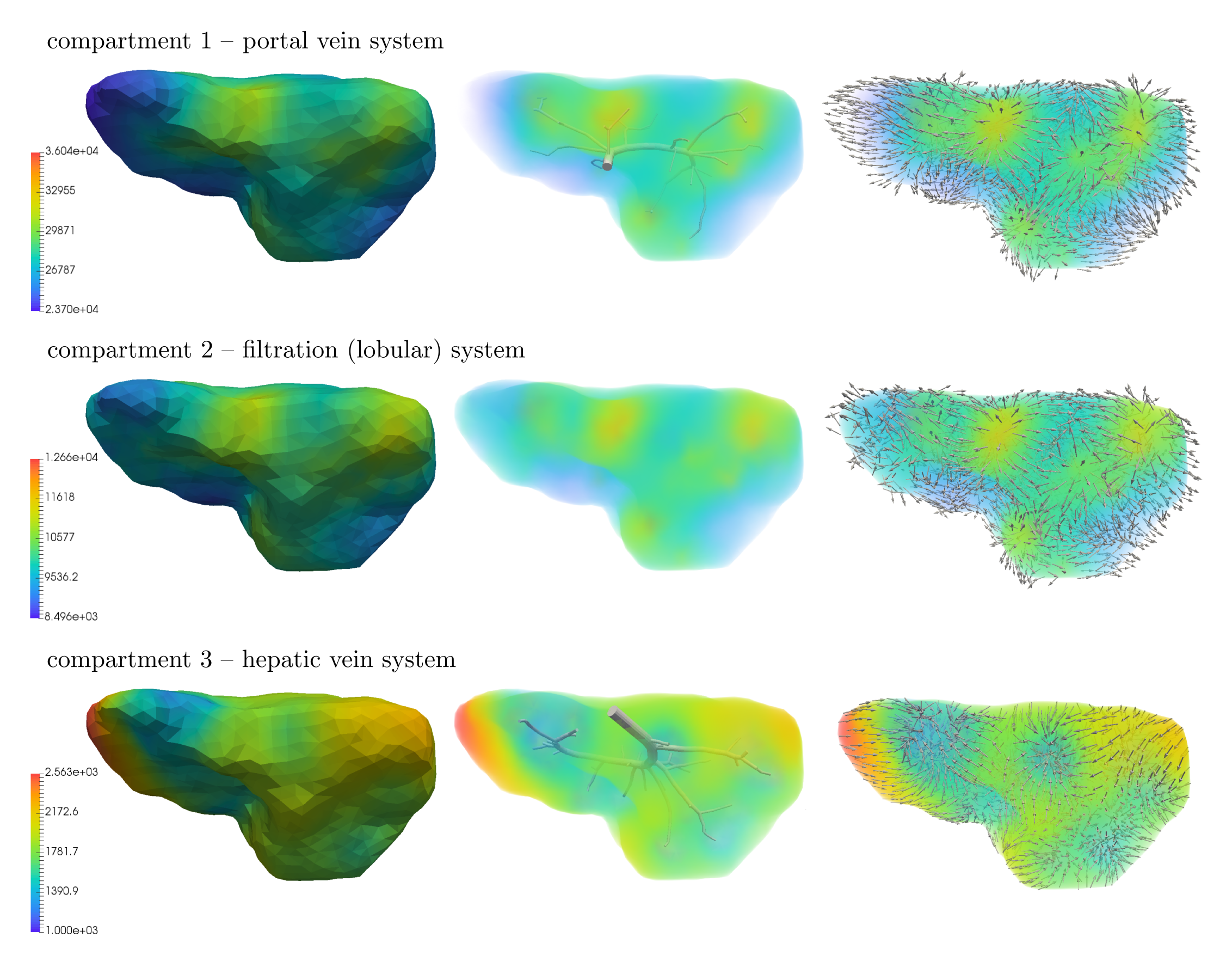}
  \caption{
    Pressures {(\it left, middle, right)}  and velocity vectors {(\it right)} in
    compartments 1--3. The upper-level vascular trees are visualized {\it
    (middle)}, they are connected to compartments 1 and 3 through 34 and 11
    junctions}
  \label{fig-liver_pressure_velocity}
\end{figure}

The 1D flow model is connected to the multicompartment model through 45
junctions, 34 of them are the sources located in compartment~1 and remaining 11
are the sinks located in compartment~3. The upper-level vascular trees, terminal
points of which correspond to the sources and sinks, are shown in
Fig.~\ref{fig-sources_sinks} and Fig.~\ref{fig-liver_pressure_velocity} {\it
(middle)}. The input blood velocity to the portal vein is taken as $v_{in} =
0.25\,{\textrm m}\cdot {\textrm s}^{-1}$ and the output blood pressure
prescribed in the hepatic vein is $p_{out} = 10^3$\,Pa.

The steady state results computed for the three compartment perfusion model
described above are presented in Fig.~\ref{fig-liver_pressure_velocity}. This
figure shows the distribution of pressure $p^i$ in each of the three
compartments, $i=1,2,3$ (1~--~portal, 2~--~filtration, 3~--~hepatic) as well as
the corresponding velocity vectors $\wb^i$, which well illustrate an interplay
between the filling and draining functionality of the portal and hepatic
compartments, respectively. From the surface and volumetric visualizations of
the computed pressure fields (\textit{left, middle}), it can be noticed that the
portal (\textit{top}) and hepatic (\textit{bottom}) pressures depend strongly on
the structure of the vascular trees, especially their upper-level hierarchies;
this is in accordance with observations made on real human liver.

\begin{figure}
  \centering
  \includegraphics[width=0.65\linewidth]{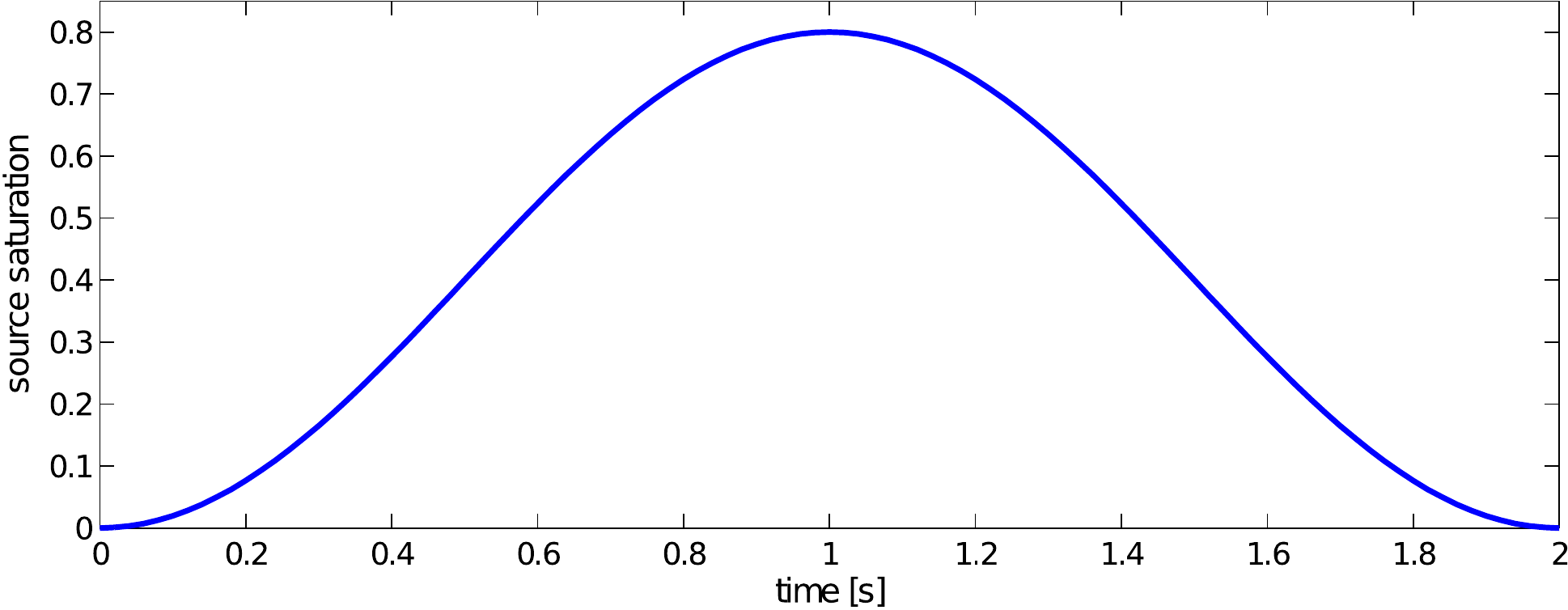}
  \caption{
    Input function $S_{\textrm{in}}(t)$ of the tracer bolus prescribed at the
    inlet of the portal vein tree}
  \label{fig-bolus}
\end{figure}

\begin{figure}
  \centering
  \includegraphics[width=0.7\linewidth]{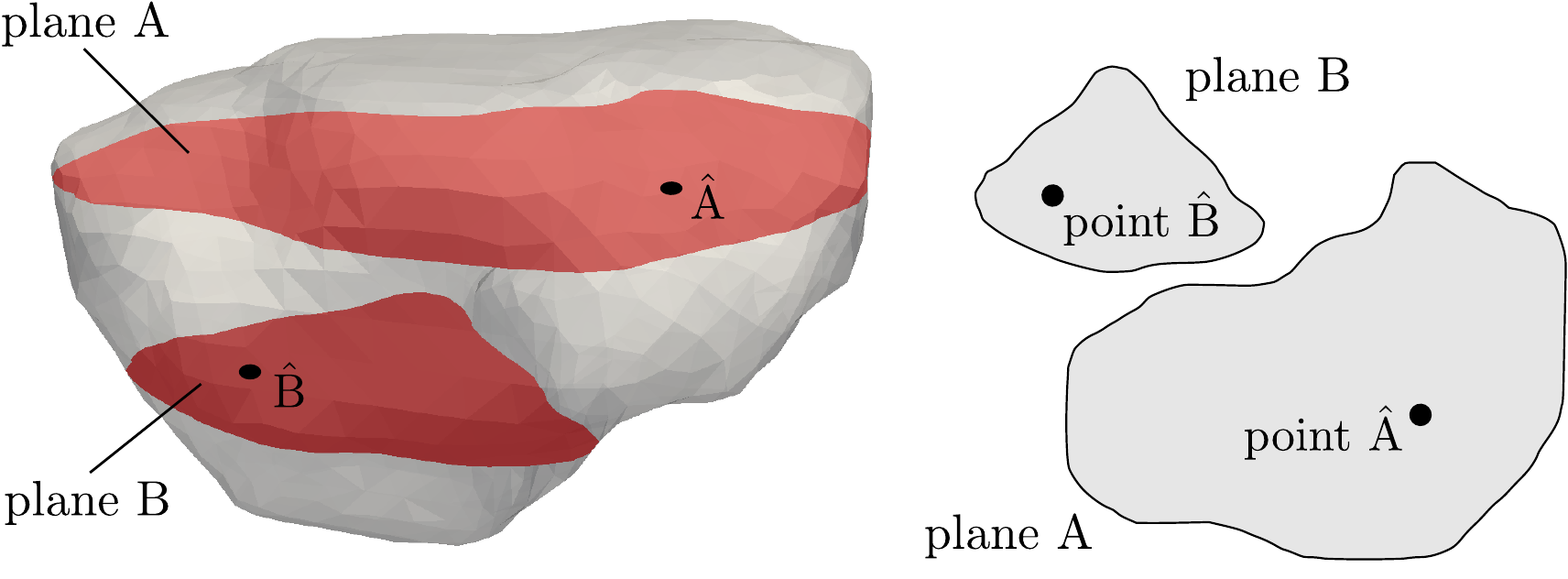}
  \caption{Two plane sections of the liver parenchyma labeled as A, B
    and points $\hat{\textrm A}$, $\hat{\textrm B}$ in these planes}
  \label{fig-liver_sec_AB}
\end{figure}

\begin{figure}
  \centering
  \includegraphics[width=0.95\linewidth]{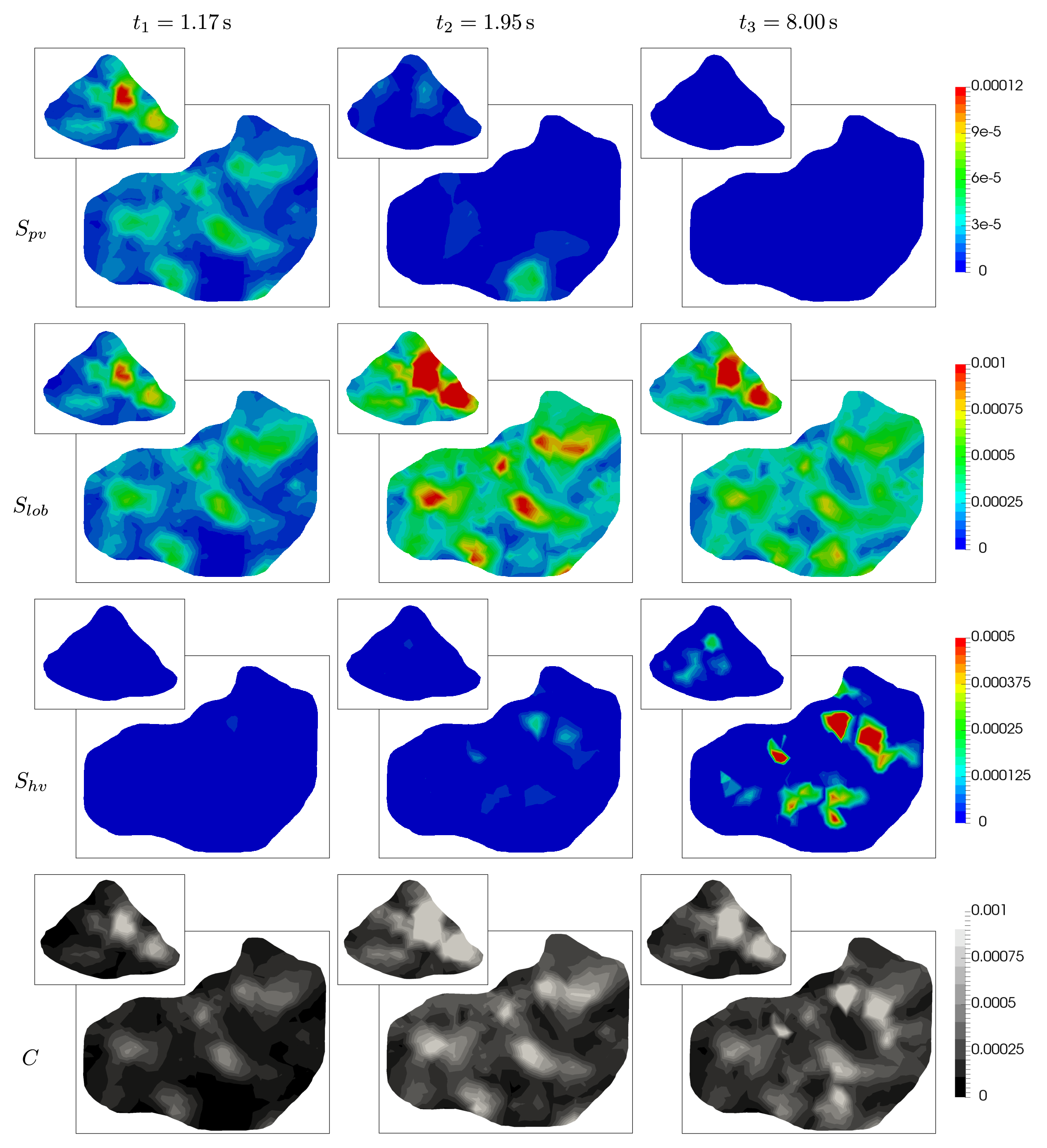}
  \caption{Distribution of portal $S_{pv}$, lobular $S_{lob}$ and
    hepatic $S_{hv}$
    saturations and total apparent concentration $C$ (\textit{from top to
    bottom}) in planes A and B of the liver at three selected time instants:
    $t_1=1.17\,\textrm{s}$, $t_2=1.95\,\textrm{s}$, and $t_3=8\,\textrm{s}$
    (\textit{from left to right})}
  \label{fig-sections-healthy}
\end{figure}

\begin{figure}
  \centering
  \includegraphics[width=0.480\linewidth]{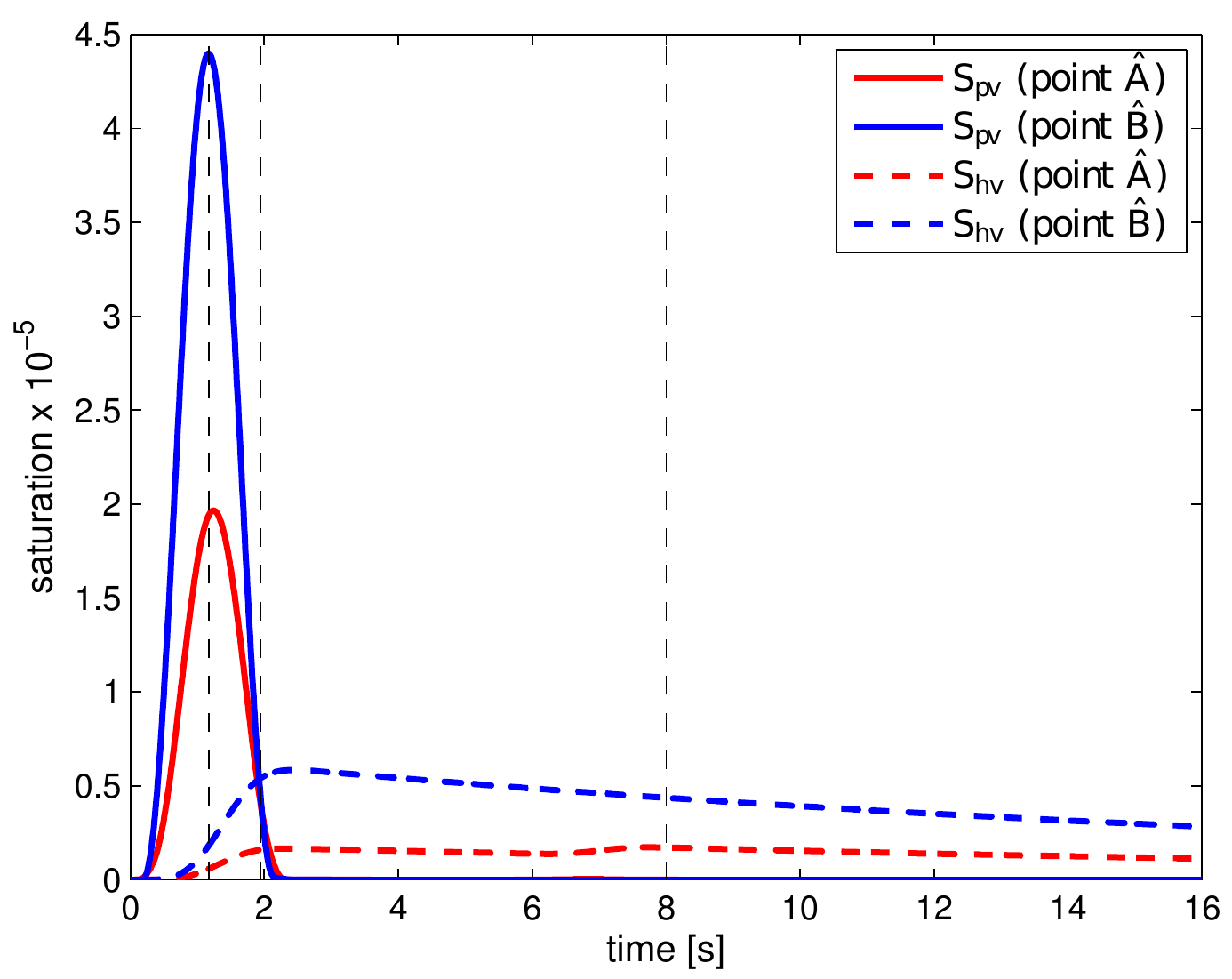}\hfill
  \includegraphics[width=0.470\linewidth]{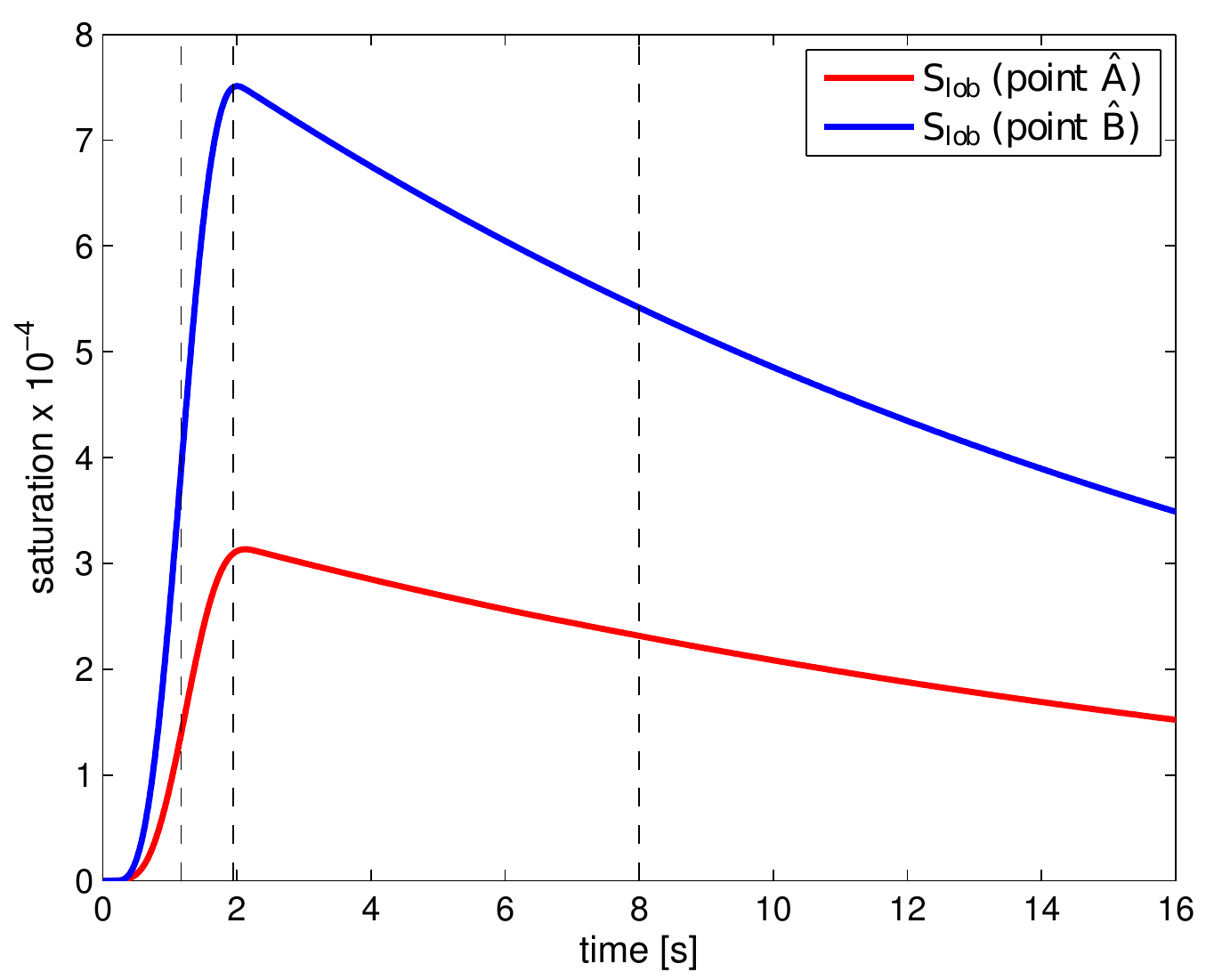}\\
  \includegraphics[width=0.470\linewidth]{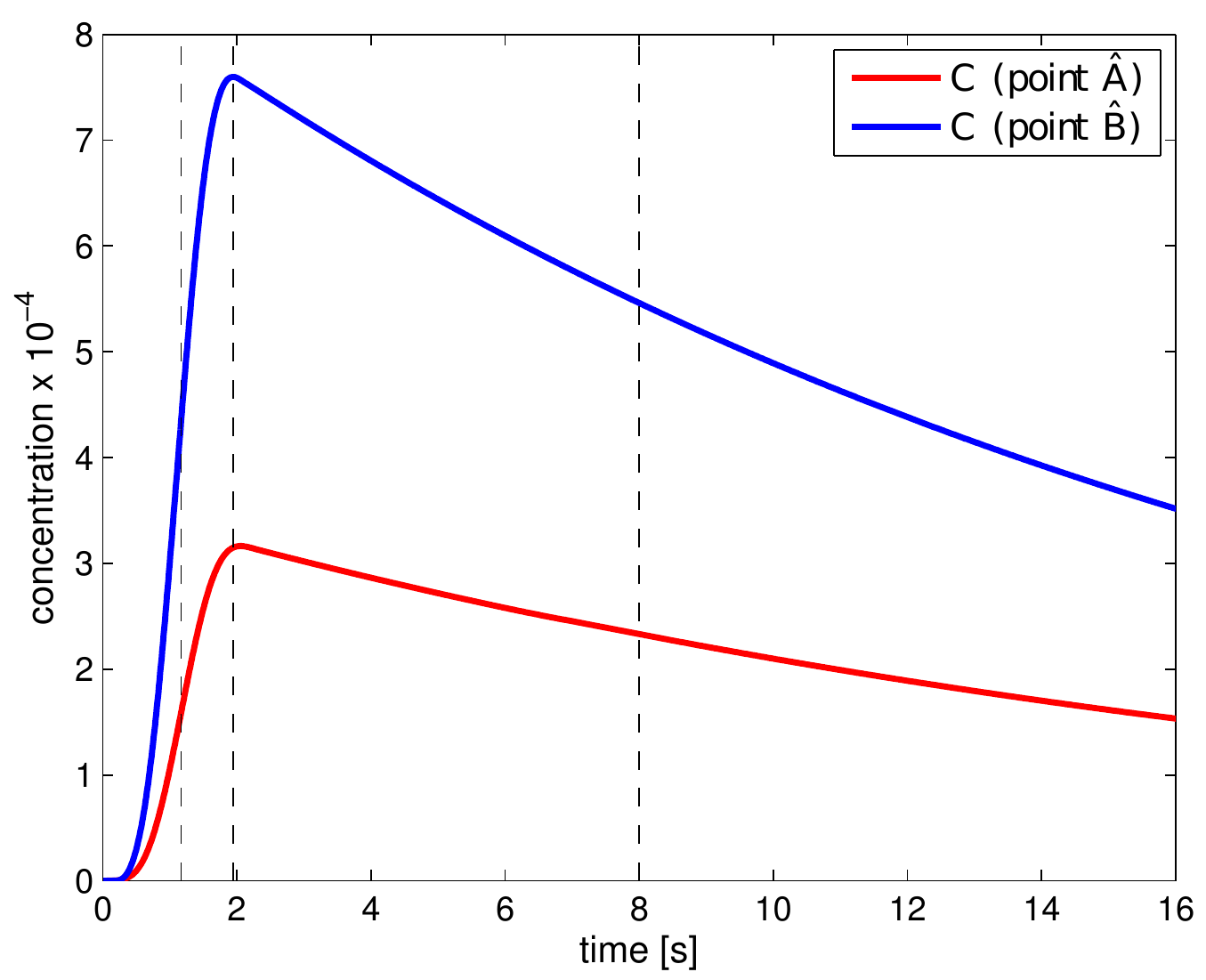}
  \caption{
    Time development of portal $S_{pv}$, lobular $S_{lob}$ and hepatic $S_{hv}$
    saturations and total concentration $C$ plotted at points $\hat{\textrm A}$
    and $\hat{\textrm B}$ shown in Fig.~\ref{fig-liver_sec_AB}. The dashed
    vertical lines in both graphs denote the three selected time instants
    $t_1=1.17\,\textrm{s}$, $t_2=1.95\,\textrm{s}$, and $t_3=8\,\textrm{s}$
    mentioned in Fig.~\ref{fig-sections-healthy}}
  \label{fig-graphs.healthy}
\end{figure}

To get a better idea on how the different perfusion properties of the three
compartments influence the final distribution of blood within the liver and also
to simulate a real dynamic perfusion test, a simulation of the contrast fluid
transport is carried out using the model presented in
Section~\ref{sec-CT}. For this purpose, a given external source saturation
$S_{\textrm{in}}(t)$ is prescribed in the form of a time bolus defined at the
root segment of the portal vein system, Fig.~\ref{fig-bolus},
\begin{equation}
S_{\textrm{in}}(t) =\left\{
\begin{array}[c]{ll}
 \bar{S}(1-\cos 2\pi\frac{t}{T}) & \textrm{for }0 \leq t\leq T,\\
 0 & \textrm{for }t>T,
\end{array}\right.
\label{eq.S-inlet}
\end{equation}
where $\bar{S}=0.4$ and $T=2\,\textrm{s}$. In Fig.~\ref{fig-sections-healthy},
the obtained portal $S_{pv}$, lobular $S_{lob}$ and hepatic $S_{hv}$ saturations
and total apparent concentration $C$ are shown in two selected sections of the
liver, positions of which are outlined in Fig.~\ref{fig-liver_sec_AB} and
denoted as planes A and B. Due to the time-dependent character of the tracer
transport simulation, the results in Fig.~\ref{fig-sections-healthy} are
visualized at three time instants ($t_1=1.17\,\textrm{s}$,
$t_2=1.95\,\textrm{s}$, and $t_3=8\,\textrm{s}$). Note that the first two time
instants reflect the time at which most of the portal $S_{pv}$ and lobular
$S_{lob}$ saturations attain their maximum, see graphs in
Fig.~\ref{fig-graphs.healthy}. The top graphs show the time development of the
portal $S_{pv}$, hepatic $S_{hv}$ and lobular $S_{lob}$ saturations at points
$\hat{\textrm{A}}$ and $\hat{\textrm{B}}$ located in planes A and B
(Fig.~\ref{fig-liver_sec_AB}), while the bottom graph depicts the evolution of
the total apparent concentration $C$ at the same points.

The saturation isocontours presented in Fig.~\ref{fig-sections-healthy} clearly
demonstrate the transport of the contrast fluid through each of the three
considered liver compartments. In other words, the most apparent saturation of
the respective compartment occurs successively---first in the portal vein system
at the time $t_1=1.17\,\textrm{s}$, followed by the lobular one at
$t_2=1.95\,\textrm{s}$ and then concluded in the hepatic vein system at
$t_3=8\,\textrm{s}$. This dynamics of the perfusion can also be recognized from
the varying grey levels of the total apparent concentration $C$ in
Fig.~\ref{fig-sections-healthy}.

To get another, more time continuous view of the tracer transport through each
of the three compartments, we refer to the graphs in
Fig.~\ref{fig-graphs.healthy}; from there it is possible to observe a relatively
fast ``filtration'' of the tracer through the portal compartment ($S_{pv}$
reaches its maximum shortly after the saturation bolus have entered the
compartment; moreover, the shape of the bolus remains almost unchanged over
time, see Fig.~\ref{fig-graphs.healthy} (\textit{top left}). On contrary, the
filtration compartment, which is characterized by a relatively high porosity
($\phi^2 = 0.15$) compared to the ones computed in the filling/draining
compartments ($\phi^1$ and $\phi^3\approx 10^{-4}$, see
Tab.~\ref{tab-perf_params}), fulfills its function as a filtration system, in
which the saturation $S_{lob}$ decreases only slowly over time, see
Fig.~\ref{fig-graphs.healthy} (\textit{top right}). As a consequence, the
perfusion properties of the filtration system influence the time development of
the hepatic saturation $S_{hv}$ which, compared to the portal saturation
$S_{pv}$, changes much more gradually depending on tracer transfer from the
filtration system.

\paragraph{Liver model with pathologically changed permeability}

\begin{figure}
  \centering
  \includegraphics[width=0.65\linewidth]{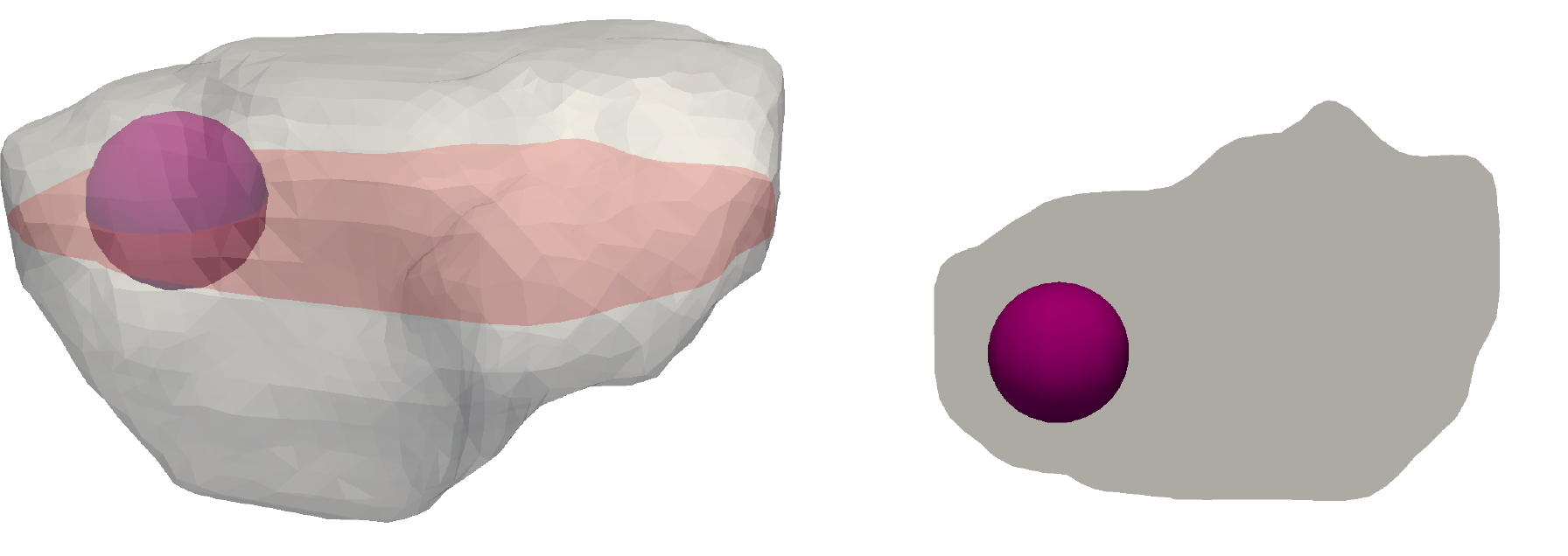}
  \caption{
    Region in the liver where the permeability in compartment 2 (the filtration
    one) is taken close to zero, this simulates a local pathology of the lobular
    structure
  }
  \label{fig-pulec}
\end{figure}

To demonstrate the applicability of the proposed model for numerical simulation
of various liver tissue pathologies, a simple illustrative example is presented.
In a small spherical region, which is depicted in Fig.~\ref{fig-pulec}, the
permeability parameter $\Kb$ of compartment 2 is locally changed to values close
to zero (due to numerical reasons arising from the solution of the Darcy
system). This change in permeability should mimic a pathology of the lobular
structure, namely a lesion.

\begin{figure}
  \centering
  \includegraphics[width=0.95\linewidth]{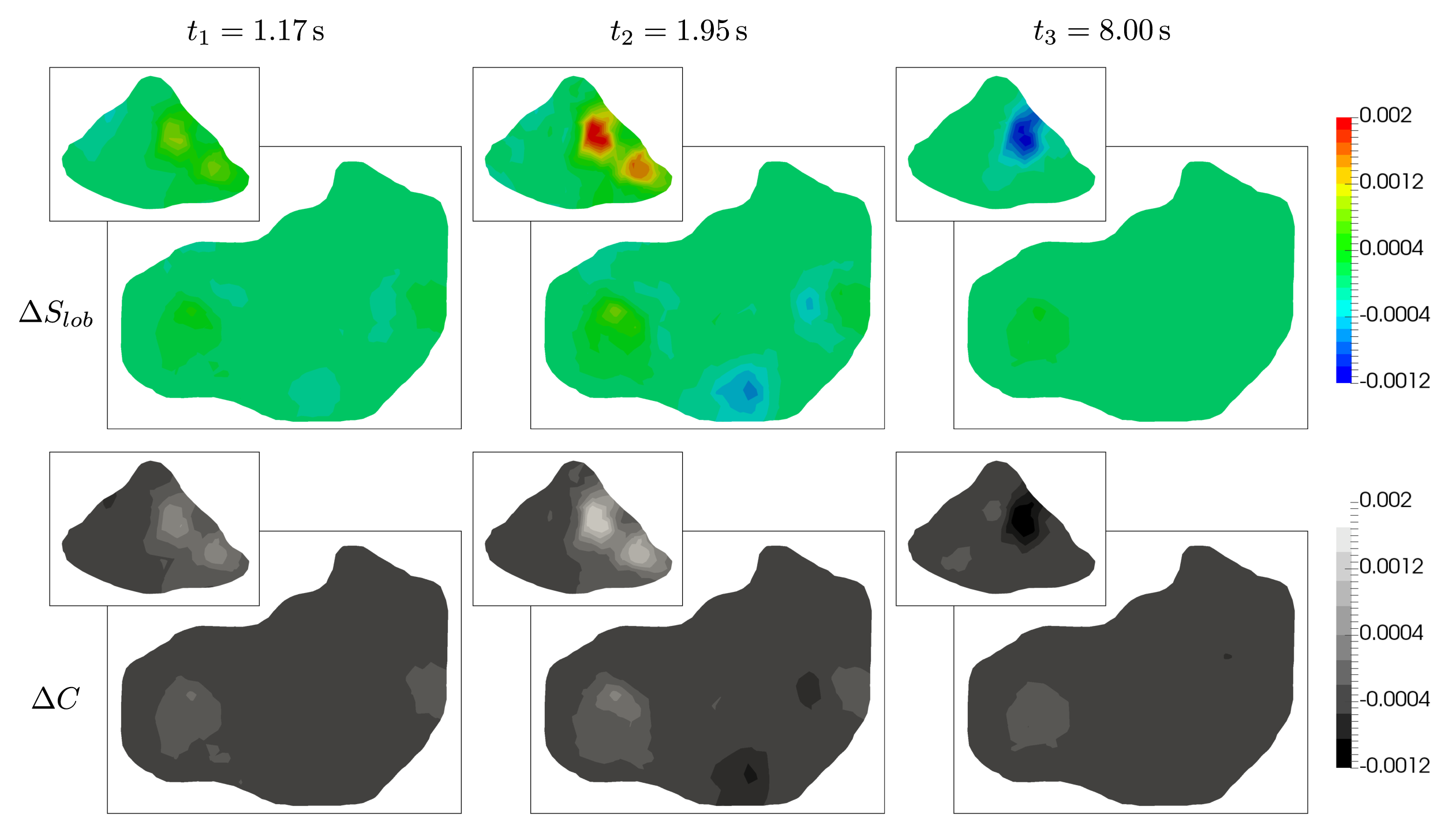}
  \caption{
    Liver model with damaged parenchyma -- distribution of $\Delta S_{pv}$ and
    $\Delta C$ (\textit{from top to bottom}) in planes A and B at three selected
    time instants $t_1=1.17\,\textrm{s}$, $t_2=1.95\,\textrm{s}$, and
    $t_3=8\,\textrm{s}$ (\textit{from left to right})}
  \label{fig-sections-damaged}
\end{figure}

To be able to quantify how the introduction of the lesion within the parenchyma
affects the overall perfusion of the liver including the transport of the
contrast fluid, the following auxiliary quantities are introduced:
\begin{equation}
 \begin{array}[c]{rcl}
  \Delta S_{pv} &=& S_{pv} - S_{pv}^{\bullet},\\
  \Delta S_{lob} &=& S_{lob} - S_{lob}^{\bullet},\\
  \Delta S_{hv} &=& S_{hv} - S_{hv}^{\bullet},\\
  \Delta C &=& C - C^{\bullet},
 \end{array}
 \label{eq-deltas}
\end{equation}
where the upper index ``$^{\bullet}$'' denotes a quantity associated with the
pathologically changed liver model. Similarly to
Fig.~\ref{fig-sections-healthy}, Fig.~\ref{fig-sections-damaged} shows the
distribution of $\Delta S_{lob}$ and $\Delta C$ in planes A and B at the three
previously introduced time instants ($t_1=1.17\,\textrm{s}$,
$t_2=1.95\,\textrm{s}$, and $t_3=8\,\textrm{s}$). From the $\Delta S_{lob}$
isocontours, it can be deduced that the lesion, which is located in the upper
part of the liver model (near plane A), affects  the perfusion and tracer
transport globally. This phenomenon can be particularly observed on plane B,
where at the beginning of the simulation ($t_2=1.95\,\textrm{s}$) the lobular
saturation $S_{lob}$  of the original model is higher than in the case of the
model with lesion (i.e., $\Delta S_{lob}>0$). The presence of the lesion within
the liver model and its influence on the overall blood perfusion is also
reflected by the distribution of $\Delta C$ in Fig.~\ref{fig-sections-damaged}
(\textit{bottom}). Here the differences in the total concentration $\Delta C$
are apparent on both planes A and B. The region with $\Delta C>0$ in the left
part of plane A is of particular importance, as it is clearly
distinguishable in all snapshots, thus, indicating the location of the
lesion.

\begin{figure}
  \centering
  \includegraphics[width=0.47\linewidth]{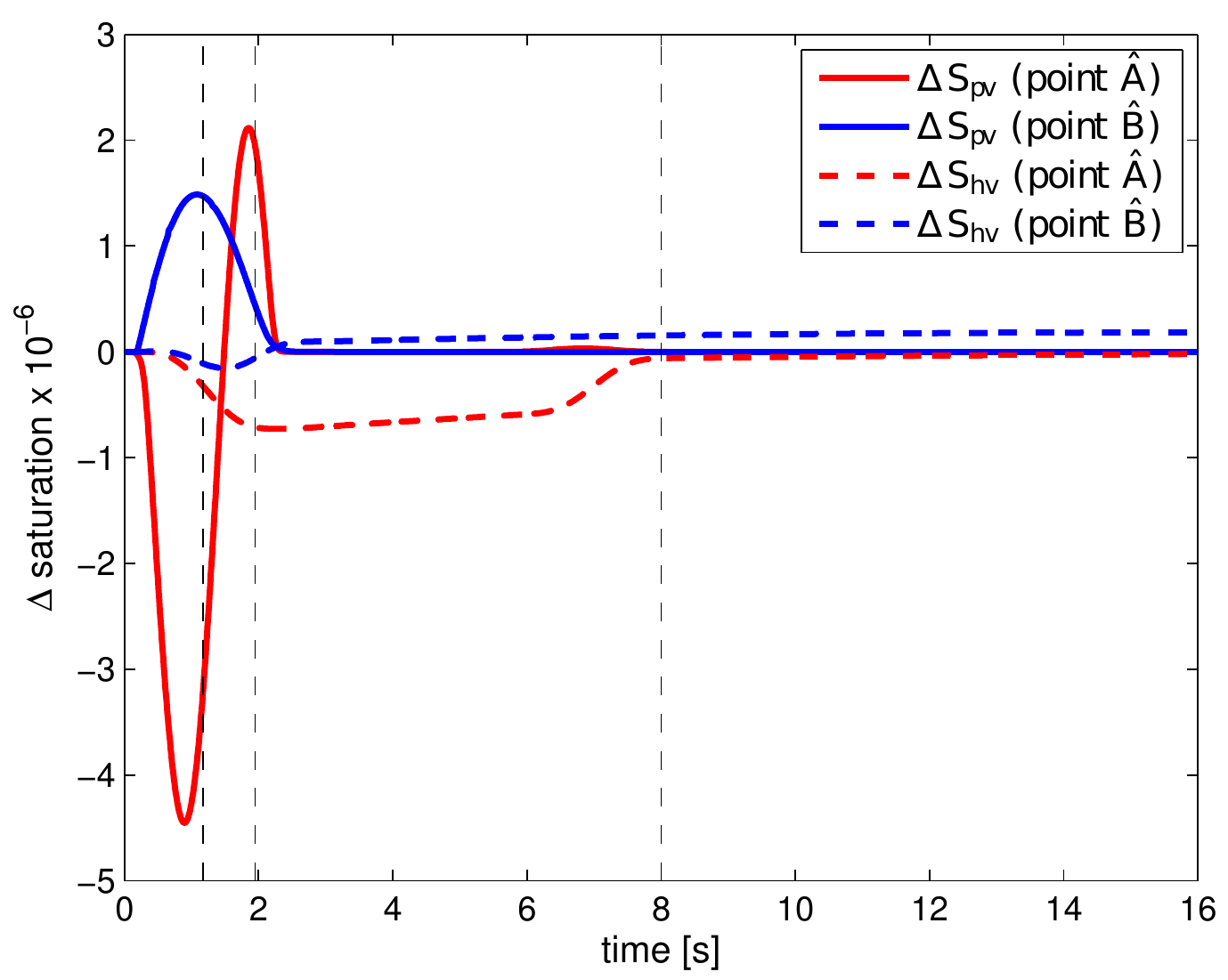}\hfill
  \includegraphics[width=0.47\linewidth]{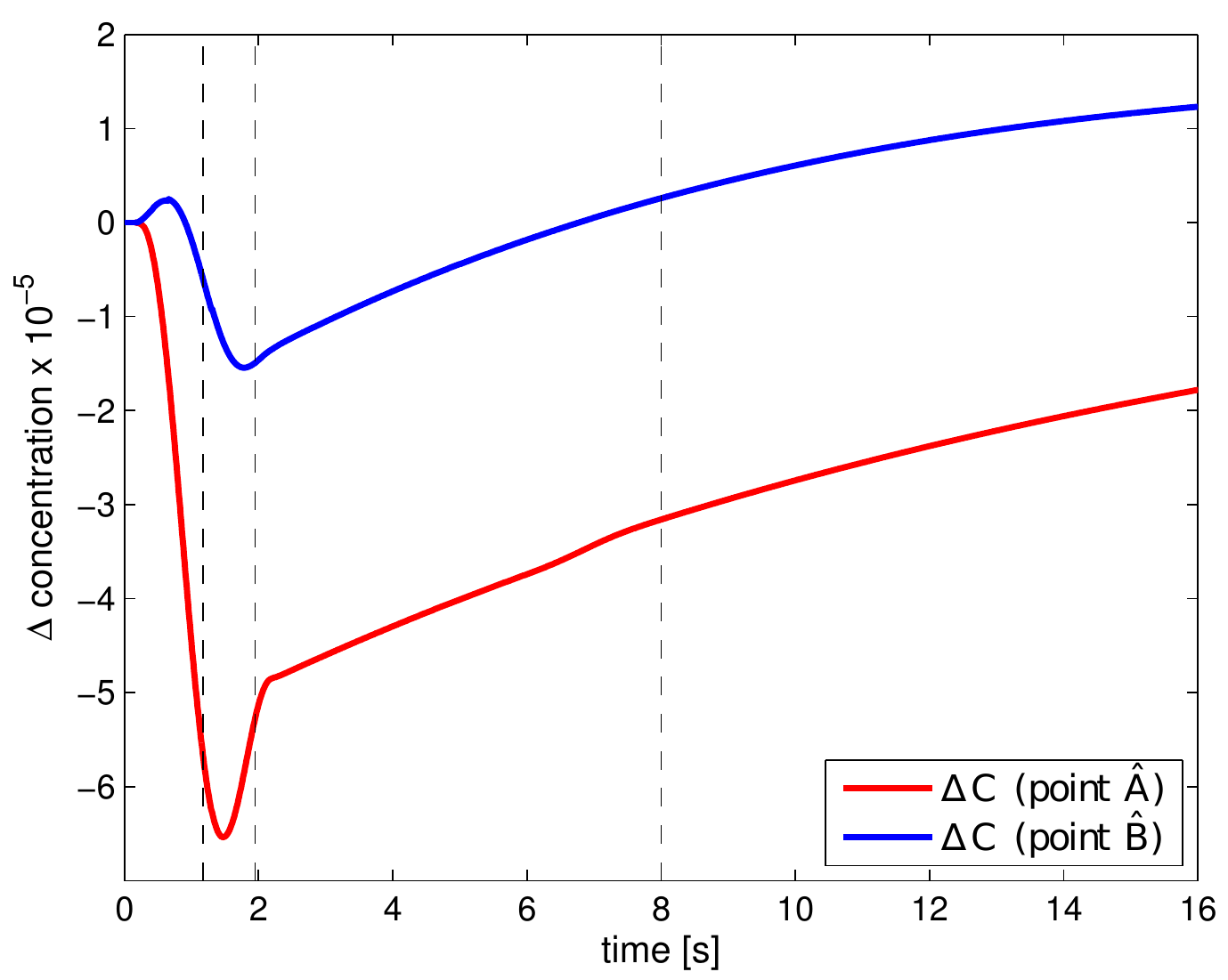}
  \caption{
    Evolution of $\Delta S_{pv}$, $\Delta S_{hv}$ (\textit{left}) and
    $\Delta C$ (\textit{right}) at points $\hat{\textrm A}$ and $\hat{\textrm
    B}$}
  \label{fig-graphs.difference}
\end{figure}

To assess the impact of the lesion on the perfusion results in a time-continuous
manner, we refer to the graphs shown in Fig.~\ref{fig-graphs.difference}, which
capture evolution of $\Delta S_{pv}$, $\Delta S_{hv}$ and $\Delta C$ in time at
the points $\hat{\textrm A}$ and $\hat{\textrm B}$ previously introduced. Note
that, in comparison with Fig.~\ref{fig-graphs.healthy}, the graph of $\Delta
S_{lob}$ is not displayed in Fig.~\ref{fig-graphs.difference}, as it would
closely resemble that of $\Delta C$. When comparing the plotted curves, two
observations are worth of nothing. First, although the perfusion anomaly
(lesion) is located only in the filtration compartment, it influences the
perfusion and, thereby, the tracer transport through the higher hierarchies
included in to the portal compartment (the difference $\Delta S_{pv}$ is not
zero from the beginning, \ie for $t < t_2$). Second, curves plotted for the
selected points $\hat{\textrm A}$ and $\hat{\textrm B}$ are dissimilar each
other, no common features are evident. This is particularly apparent in the case
of $\Delta S_{pv}$, which for both points exhibit completely different
behaviour: On one hand, at point $\hat{\textrm A}$ (red solid line in
Fig.~\ref{fig-graphs.healthy} (\textit{left})), the graph is characterized by a
delayed maximum portal saturation in the undamaged liver model (fast change of
$\Delta S_{pv}$ from negative to positive values within a short time interval).
On the other hand, the time evolution of $\Delta S_{pv}$ at point $\hat{\textrm
B}$ (blue solid line in Fig.~\ref{fig-graphs.healthy} (\textit{left})) shows no
time delay, only higher portal saturation compared to the damaged liver model is
apparent ($\Delta S_{pv}>0$ during the whole simulation). Similar observations
can be made in the case of the $\Delta S_{hv}$ and $\Delta C$ curves.

\section{Conclusions}\label{sec-Concl}

In the paper, we proposed the multicompartment model of the tissue blood
perfusion and of the superimposed time-space distribution of the contrast fluid.
This latter feature enables to establish a modeling feedback which is necessary
to tune the model parameters according to information available due to standard
clinical CT examination combined with essential flow parameters and basic
structural and morphological data of a specific perfused tissue. Although the
model is intended and currently being developed for modeling the liver perfusion, it
can be adapted for other applications, namely the cerebral perfusion; this issue
is in our focus for future work.

\subsection{Summary of the proposed modeling approach}

The perfusion model has been derived on a very simple idea of decomposing the
perfusion tree into a number of sets -- the compartments -- which are
hierarchically organized. The flow within each compartment $i$ is governed by
the Darcy law involving permeability tensors $\Kb^i$. Cross links between two
``communicating'' compartments $i$ and $j$ is respected locally by scalar
perfusion coefficients $G_i^j$ which are associated with  cuts of the perfusion
tree at those bifurcations separating the two different hierarchies. These
features represent a substantial step towards an anatomically parametrized
porous perfusion model. However, the key issue is the quantification of the
model parameters. If an exact geometry of the perfusion tree is known, as in our
case, when using the artificially generated trees,  the permeabilities $\Kb^i$
can be determined using the geometrical data by a representative volume
averaging following the approaches suggested in
\cite{huyghe_campen_1995:ALL,vankan-huyghe_1997} and followed in
\cite{Michler2013}. Following this work, a procedure has been proposed to
determine also the perfusion coefficients $G_i^j$ which, however, would require
resolving the whole ``non-reduced'' problem of flow on all vessel segments of
the tree. Therefore, in our treatment, we simplified such a complex procedure to
obtain an approximate, but tractable computation of the perfusion coefficients.

Flow on the uppermost hierarchies of the perfusion tree is described by the
Bernoulli type model on 1D branching network; its coupling with the 3D continuum
multicompartment model is by means of distributed point sources, or sinks. An
iterative algorithm is based on commuting the 1D flow solvers associated with
the portal and hepatic vein trees, and the 3D perfusion solver, whereby the
inlet velocity at the portal vein is being increased gradually from zero until a
steady state is reached.

As the second contribution of the paper, we propose the model describing the
contrast fluid (the tracer) dynamic transport at all hierarchies of the
perfusion trees. Superposition of the local saturation  of the tracer in all
compartments yields the local concentration, often called the tissue density
which can be measured by the perfusion CT examination. On one hand, due to this
option, it is possible to use the patient-specific CT images to tune the model
parameters. On the other hand, this computational tool will allow for a deeper
analysis of CT scans and a more accurate localization and assessment of possible
liver pathology.

\subsection{Discussion and future development of the model}

To bring the modeling approach to its practical application with valuable
outputs for clinical practice, there are several important issues to be pursued
in our future work. The main purpose of using the multicompartment continuum
model to describe approximately the flow on complex branching vessel networks,
like trees, is to avoid direct flow simulations  which may be prohibitively
expensive and even non-feasible to provide real time solutions.

To verify the model performance, we need to check how the tree decomposition
into varying number of compartments influences the modeling error. Such a
numerical experiment requires a reference model which would provide a
sufficiently fine and robust resolution of flow on all the vascular tree
branches. For this purpose, we shall consider the 1D flow on the perfusion tree
described by the Bernoulli-type model. This reference model will also be used to
set the parameters $G_i^j$ for a particular split of the tree.

For practical use of the model, its parameters cannot be set appropriately using
the direct calculations: the reason is twofold. If real data is used, the tree
cannot be identified to a sufficient resolution of small vessels using the
imaging techniques which are currently available. Moreover, even if the tree is
known, as in our numerical examples, a cumbersome direct calculation of the flow
would be necessary to set the parameters $G_i^j$. Therefore, we intend to
identify the model parameters by solving an inverse problem. In
\cite{Rohan-vipimage2015}, we suggested a formulation of the nonlinear least
square problem which is based on the discrepancy between the local contrast
fluid concentration resolved by the model, and the corresponding data provided
by the  CT examination. This approach seems to be the only way of tuning the
model. However, such a treatment requires further improvements of the model
itself to capture more accurately the flow at the lowermost level of the
parenchyma and also to account for dispersion of the contrast fluid during its
transport through the compartments.

We are currently working on a homogenized model \citep{RTL-CC2015} of the flow
on the capillary and pre-capillary level which should be integrated in the
multicompartment model. In liver, the lobular level of the vasculature can be
approximated by a periodic structure featured by the double porosity in the
context of the vertex and the central venules connected by the hepatic
capillaries of the sinusoids. Another issue is certainly related to
the deformation of the parenchyma
\citep{Cimrman2007,rohan-cimrman-perfusionIJMCE2010,RL-CST-2014}. In a longer
perspective, this phenomenon must be treated in the model to capture processes
of the tissue remodeling.

  %
\acknowledgements{
  This research is supported by 
  the project LO\,1506 of the Czech Ministry of
  Education,Youth and Sports.}
 %


\bibliographystyle{spbasic}      

\bibliography{biblio-liver}

\end{document}